\documentclass[eqsecnum,noshowpacs,showkeys,nofootinbib,aps]{revtex4}
\usepackage{graphicx}
\usepackage{amsmath,latexsym}
\usepackage{amssymb}

\renewcommand{\theequation}{\arabic{equation}}
\def\beq{\begin{equation}}
\def\eeq{\end{equation}}
\def\bea{\begin{eqnarray}}
\def\eea{\end{eqnarray}}
\def\nn{\nonumber}

\def\pa{\partial}

\def\na{\nabla}

\begin{document}
\title{New algorithm of measuring gravitational wave radiation from rotating binary system}
\author{Soon-Tae Hong}
\email{galaxy.mass@gmail.com}
\affiliation{Center for Quantum Spacetime and Department of Physics, Sogang University, Seoul 04107, Korea}
%\date{August 29, 2003}
\date{\today}
\begin{abstract}
In order to investigate the gravitational wave (GW) radiation, without resorting to the traceless transverse gauge approach to 
the GW formalism of the linearized general relativity, we formulate the so-called modified 
linearized general relativity (MLGR). As an application of the MLGR, we construct a novel paradigm of measuring the GW radiation from 
a binary system of compact objects, to theoretically interpret its phenomenology. To accomplish this, we formulate 
the mass scalar and mass vector potentials for the merging binary compact objects, from which we construct 
the mass magnetic field in addition to the mass electric one which also includes 
the mass vector potential effect. Next, defining the mass Poyinting vector in terms of 
the mass electric and mass magnetic fields in the MLGR, we find the GW radiation intensity profile possessing 
a prolate ellipsoid geometry due to the merging binary compact objects source. At a given radial distance from the binary compact objects, 
the GW radiation intensity on the revolution axis of the binary compact objects is shown to be twice that on 
the equatorial plane. Moreover, we explicitly obtain the total radiation power of the GW, which has the 
same characteristic as that of the electromagnetic wave in the rotating charge electric dipole moment. 
We also find that, in no distorting limit of the merging binary compact objects, the compact objects in the MLGR
do not yield the total GW radiation power, consistent with the result of the traceless transverse gauge algorithm in the linearized general relativity. 
\end{abstract}
%\pacs{12.39.Dc, 14.20.Dh, 45.20.Jj} 
\maketitle

%%%%%%%%%%%%%%%%%%%%%%%%%%%%%%%%%%%%%%%%%%%%%%%%%%%%%%%%%%%%%%%%%%%%%%%%
\section{Introduction}
\setcounter{equation}{0}
\renewcommand{\theequation}{\arabic{section}.\arabic{equation}}
%%%%%%%%%%%%%%%%%%%%%%%%%%%%%%%%%%%%%%%%%%%%%%%%%%%%%%%%%%%%%%%%%%%%%%%%

The linearized general relativity (LGR)~\cite{wald84,hobson06,carroll04} 
has been investigated to describe the gravitational wave (GW) radiation from a binary system of compact objects 
(namely stars or black holes) for instance. To be specific, in order to study the physical phenomenology in the LGR, we need to assume that the 
gravity fields are weak so that the gravitational field equations can be linearized. 
Next it is well known that, in vacuum of the LGR, the binary compact objects oscillating in time emit spin-two gravitons which are 
quanta of the GW. Mathematically the spin-two gravitons originate from the second rank tensor 
associated with the curved spacetime metric $g_{\alpha\beta}$~\cite{pauli39}. In order to construct 
the GW which is transverse to the GW propagating direction and yields the spin-two graviton, we have exploited 
the traceless transverse (TT) gauge~\cite{wald84,hobson06,carroll04}. In contrast, the spin-one photons 
which are quanta of the electromagnetic (EM) wave in vacuum are delineated in terms of the vectorial quantities of the charge 
scalar and charge vector potentials defined in the flat spacetime. 

On the other hand, since the first GW was detected by Advanced LIGO~\cite{aasi15}, there have been lots of progresses in an observational astrophysics. 
In particular, the third generation network of ground-based detectors~\cite{reitze19} is currently proposed to obtain improved sensitivities, 
compared to those of the Advanced LIGO detector. Recently, the GW from neutron star and black hole has been observed~\cite{ligo21}, and 
the galaxy-based observatories have been proposed~\cite{dan23}. In order to constrain cosmology, a theoretical probe of cosmology which exploits 
population level properties of lensed GW detections has been also proposed ~\cite{jana23}. Next, since the gamma ray burst (GRB) discovery was 
published~\cite{strong73}, to elucidate the GRB associated with the magnetar (namely highly-magnetized neutron star) 
for instance~\cite{hurley21}, we have observed many theoretical models including the Dirac type relativistic 
massive photon model~\cite{hong22qm}, where the anti-photon corresponding to the negative energy 
solution is proposed as a candidate for an intense radiation flare of the GRB. The GW detector network has observed the 
GW merger event originating from binary neutron star GW170817 in 2017~\cite{abbott2017a,grb1,goldstein2017}. 
Note that the GW170817 merger event has been known to be consistent with the GRB of GRB170817A~\cite{grb1,goldstein2017}. 
To be specific, the GW is supposed to be produced from neutron star oscillations 
related with magnetar giant flares and the corresponding GRB~\cite{kaspi2017}. 

In this paper, we will propose a new theoretical paradigm of the modified linearized general relativity (MLGR), in which we will formulate 
the Green retarded~\cite{wald84,hobson06} mass scalar and mass vector potentials of a binary system of compact objects. Using 
these potentials in the MLGR, we will next construct the mass electric and mass magnetic fields. In particular, in order to formulate the 
non-vanishing mass magnetic field, we will introduce the distortion of the spirally merging binary compact objects during one cycle 
of their revolution. Making use of the mass electric and mass magnetic fields, we will find the mass Poyinting vector which is exploited to construct 
a theoretical scheme of measuring the GW radiation from a binary system of compact objects. Moreover, introducing the 
TT gauge, we will study the GW of the binary compact objects in the LGR.

In Section 2 we will formulate the LGR and MLGR algorithms. In Section 3, as an application of the MLGR, we will investigate 
the phenomenology of the merging binary compact objects. To do this, we will explicitly formulate the 
physical quantities such as the mass scalar and mass vector potentials, and the mass Poyinting vector. We will also study the TT 
gauge in the LGR. Section 4 includes conclusions. In Appendix A, we will briefly study the EM 
radiation from the rotating charge electric dipole moment.
In Appendix B, we will briefly comment on the TT gauge formalism for the gravitation radiation from the rotating mass compact objects. 
In Appendix C, we will also address comments on the formalism of massive spin-one graviton.

%%%%%%%%%%%%%%%%%%%%%%%%%%%%%%%%%%%%%%%%%%%%%%%%%%%%%%%%%%%%%%%%%%%%%%%%%%
\section{Set up of LGR and MLGR algorithms}
%setcounter{equation}{0}
\renewcommand{\theequation}{\arabic{section}.\arabic{equation}}
\label{setupgravitosection}
%%%%%%%%%%%%%%%%%%%%%%%%%%%%%%%%%%%%%%%%%%%%%%%%%%%%%%%%%%%%%%%%%%%%%%%%%%

%%%%%%%%%%%%%%%%%%%%%%%%%%%%%%%%%%%%%%%%%%%%%%%%%%%%%%%%%%%%%%%%%%%%%%%%%%
\subsection{LGR algorithm}
%setcounter{equation}{0}
%\renewcommand{\theequation}{\arabic{section}.\arabic{equation}}
%\label{setupgravitosection1}
%%%%%%%%%%%%%%%%%%%%%%%%%%%%%%%%%%%%%%%%%%%%%%%%%%%%%%%%%%%%%%%%%%%%%%%%%%

Now we will study the LGR~\cite{wald84,hobson06,carroll04}. To accomplish this, we first assume 
that the deviation $h_{\alpha\beta}$ of the spacetime metric $g_{\alpha\beta}$ from a flat metric 
$\eta_{\alpha\beta}$ is small: $g_{\alpha\beta}=\eta_{\alpha\beta}+h_{\alpha\beta}$. 
After some algebra we find the linearized Einstein equation~\cite{wald84}
\beq
-\frac{1}{2}\square \bar{h}_{\alpha\beta}+\pa^{\gamma}\pa_{(\alpha}\bar{h}_{\beta)\gamma}-\frac{1}{2}\eta_{\alpha\beta}
\pa^{\gamma}\pa^{\delta}\bar{h}_{\gamma\delta}=\frac{8\pi G}{c^{4}} T_{\alpha\beta},
\label{lineqm}
\eeq
where $\square=\pa^{\beta}\pa_{\beta}$ with the metric $(-1,+1,+1,+1)$, and $T_{\alpha\beta}$ is the energy stress tensor. 
Here the trace reversed perturbation $\bar{h}_{\alpha\beta}$ is given by $\bar{h}_{\alpha\beta}=h_{\alpha\beta}
-\frac{1}{2}\eta_{\alpha\beta}h$ and $h=\eta^{\alpha\beta}h_{\alpha\beta}$. 
Exploiting (\ref{lineqm}) and the Lorentz gauge condition (LGC) in the gravity
\beq
\pa^{\beta}\bar{h}_{\alpha\beta}=0,
\label{pabarh0}
\eeq
we arrive at
\beq
\square \bar{h}_{\alpha\beta}=-\frac{16\pi G}{c^{4}} T_{\alpha\beta}.
\label{squarebarh}
\eeq

Next we use the fact that the LGC in (\ref{pabarh0}) is preserved by the gauge transformation 
\beq
h_{\alpha\beta}\rightarrow h_{\alpha\beta}^{\prime}= h_{\alpha\beta}+\pa_{\alpha}\xi_{\beta}+\pa_{\beta}\xi_{\alpha},
\label{harrowhp}
\eeq
if the following restricted gauge freedom condition is exploited~\cite{wald84} 
\beq
\square \xi_{\alpha}=0.
\label{squarexi}
\eeq
In other words, using the relation 
\beq
\pa^{\beta}\bar{h}_{\alpha\beta}\rightarrow \pa^{\beta}\bar{h}^{\prime}_{\alpha\beta}=\pa^{\beta}\bar{h}_{\alpha\beta}
+\square \xi_{\alpha}.
\label{lorentz22} 
\eeq
we find that, to preserve the LGC of $\pa^{\beta}\bar{h}^{\prime}_{\alpha\beta}=0$ in (\ref{lorentz22}) in the LGR, we need to ensure that 
$\square \xi_{\alpha}=0$ as shown in (\ref{squarexi}). 
 
Now we address brief comments on the radiation gauge, and the corresponding initial value problem and dynamics 
of the massive spin-two graviton~\cite{wald84,hobson06,carroll04}. 
First, in a source free region possessing $T_{\alpha\beta}=0$ in the LGR, (\ref{squarebarh}) becomes 
\beq
\square \bar{h}_{\alpha\beta}=0.
\label{squarebarh22}
\eeq
Next we use the radiation gauge~\cite{wald84,hobson06}
\beq
h=0,~~~h_{0i}=0~(i=1,2,3),
\label{hohoii}
\eeq
to yield~\cite{wald84,hobson06}
\beq
\bar{h}_{\alpha\beta}=h_{\alpha\beta},   
\label{habhab}
\eeq
and 
\beq
h_{00}=0.
\label{hoo22}
\eeq 
To find the radiation gauge in (\ref{hohoii}), on the initial surface 
$t=t_{0}$ we need to exploit (\ref{harrowhp}). For instance, the conditions $h^{\prime}=0$ and $h^{\prime}_{0i}=0$ yield
\bea  
-h&=&2\left(-\frac{\pa \xi_{0}}{\pa t}+\na\cdot\vec{\xi}\right),\nn\\
-h_{0i}&=&\frac{\xi_{i}}{\pa t}+\frac{\xi_{0}}{\pa x^{i}},
\label{cond2}
\eea
where $\xi_{\alpha}=(\xi_{0},\xi_{i})$ $(\alpha=0,1,2,3)$. Next the first time derivatives of (\ref{cond2}) produce
\bea  
-\frac{\pa h}{\pa t}&=&2\left(-\nabla^{2}\xi_{0}+\nabla\cdot\frac{\pa\vec{\xi}}{\pa t}\right),\nn\\
-\frac{\pa h_{0i}}{\pa t}&=&\nabla^{2}\xi_{i}+\frac{\pa}{\pa x^{i}}\left(\frac{\xi_{0}}{\pa t}\right).
\label{cond22}
\eea
Note that exploiting (\ref{cond2}) and (\ref{cond22}) we obtain the inertial values of $\xi_{\alpha}$ $(\alpha=0,1,2,3)$ and 
their first time derivatives, and thus we define $\xi_{\alpha}$ to be the solution of (\ref{squarexi}) with these initial data~\cite{wald84}. 
 
%-------------------to insert 11
%\textcolor{blue}{
Second, exploiting $\bar{h}_{\alpha\beta}=h_{\alpha\beta}$ in (\ref{habhab}) related with the radiation gauge in (\ref{hohoii}) the source free linearized Einstein equation produces 
plane GW solution of the form~\cite{wald84,hobson06,carroll04}
\beq
h_{\alpha\beta}=H_{\alpha\beta}e^{ik_{\mu}x^{\mu}},
\label{planewavegra}
\eeq
where $H_{\alpha\beta}$ is a constant symmetric second rank tensor field and $k^{\mu}$ is the GW four-vector. 
Note that the symmetric tensor $H_{\alpha\beta}(=H_{\beta\alpha})$ has ten 
different components. However the four LGCs in (\ref{pabarh0}) reduce the number of independent degrees of freedom (DOF) to six. Moreover the four gauge transformation conditions 
in (\ref{squarexi}) reduce the six independent components to two DOF which implies that we find two possible polarizations for the spin-two plane 
GW in (\ref{planewavegra})~\cite{wald84,hobson06}.
%}

%-------------------to insert 12
%\textcolor{blue}{
Third, we study the spin-two plane GW in the vacuum having $T_{\alpha\beta}=0$. Inserting 
$h_{\alpha\beta}$ in (\ref{planewavegra}) into (\ref{squarebarh22}), we find that 
the GW four-vector $k^{\mu}$ satisfies
\beq
k_{\mu}k^{\mu}=0,
\label{kmukmu0}
\eeq
to yield, for the plane GW traveling in the $z$ direction,
\beq
k^{\mu}=(k,0,0,k).
\label{kkk00}
\eeq
Next, combing $h_{\alpha\beta}$ in (\ref{planewavegra}) and the LGC in (\ref{pabarh0}), we obtain the transversality condition 
for the symmetric second rank tensor $H_{\mu\nu}$ 
\beq
k^{\mu}H_{\mu\nu}=0.
\label{kmukmu1}
\eeq
The aspects associated with (\ref{kmukmu0})--(\ref{kmukmu1}) imply that a massless spin-two field 
propagates with speed of light in flat spacetime~\cite{pauli39} and its quantized objects are spin-two gravitons 
possessing the two transverse polarization DOF.
Inserting the GW solution in (\ref{planewavegra}) into (\ref{hohoii}) and (\ref{hoo22}) we arrive at~\cite{wald84}
\beq
\eta^{\mu\nu}H_{\mu\nu}=0,~~~H_{0\mu}=0,~(\mu=0,1,2,3).
\label{2identities}
\eeq
%}
%-------------------

Fourth, similar to the case of the massive photon~\cite{hong22qm}, 
we find the relation for the {\it massive} spin-two graviton
\beq
k^{\mu}H_{\mu\nu}\neq 0,
\label{kalphaneq}
\eeq
since for the massive spin-two graviton we have longitudinal component in addition to transverse ones, 
as in the case of the phonon associated with massive particle lattice vibrations~\cite{phonon}. In other words, the massive 
spin-two graviton has three components to yield three DOF. Note that, as 
in the case of the GRB in the Dirac type relativistic massive photon model~\cite{hong22qm}, the massive spin-two graviton could 
be interpreted as a spin-two GRB-like particle.

%%%%%%%%%%%%%%%%%%%%%%%%%%%%%%%%%%%%%%%%%%%%%%%%%%%%%%%%%%%%%%%%%%%%%%%%%%
\subsection{MLGR algorithm}
%setcounter{equation}{0}
%\renewcommand{\theequation}{\arabic{section}.\arabic{equation}}
%\label{setupgravitosection1}
%%%%%%%%%%%%%%%%%%%%%%%%%%%%%%%%%%%%%%%%%%%%%%%%%%%%%%%%%%%%%%%%%%%%%%%%%%

Now, we will formulate the MLGR which includes explicitly the mass current DOF. 
As a model appropriate for the GW radiation from rotating binary system, we propose the MLGR where 
we have the non-vanishing $\bar{h}^{00}$ and $\bar{h}^{0i}$ $(i=1,2,3)$, together with
\beq
\bar{h}^{ij}=0.
\label{hijmlge}
\eeq 
Note that the TT gauge of the LGR related with the GW radiation will be briefly discussed in the ensuing section and Appendix B.

Next we define in the MLGR the time-time component of the trace reversed perturbation $\bar{h}^{00}$ and the 
corresponding energy stress tensor $T^{00}$ as
\beq
\bar{h}^{00}=-\frac{4}{c}A_{M}^{0},~~~T^{00}=c^{2}\rho_{M},
\label{barht0}
\eeq
where $\rho_{M}$ is the mass density. Here the subscript $M$ denotes the mass related quantities and will be 
applied to the gravitational interaction physical quantities from now on. Inserting (\ref{barht0}) into (\ref{squarebarh}) yields
\beq
\square A_{M}^{0}=\frac{4\pi G}{c}\rho_{M},
\label{naphi0}
\eeq
implying that we have the dynamical DOF in the MLGR. 

Making use of the definition $\phi_{M}\equiv cA_{M}^{0}$ and (\ref{barht0}), we find 
\beq
\bar{h}^{00}=-\frac{4}{c^{2}}\phi_{M},
\label{h00phim}
\eeq
and, in the static limit, (\ref{naphi0}) produces
\beq
\na^{2} \phi_{M}=4\pi G\rho_{M},
\label{newtongr}
\eeq
which is in agreement with the Newtonian gravity~\cite{goldstein80,wald84}. 
In contrast, in the TT gauge we find $\bar{h}^{00}_{TT}=0$ in (\ref{standardlgr}) implying that this gauge is not simultaneously applicable to 
the Newtonian gravity.

Next in the MLGR we define the time-space components of the trace reversed perturbation $\bar{h}^{0i}$ and the energy stress tensor $T^{0i}$ as
\beq
\bar{h}^{0i}=-\frac{4}{c}A^{i}_{M},~~~T^{0i}=cJ^{i}_{M},
\label{barht2}
\eeq
where $J^{i}_{M}\equiv \rho_{M}v^{i}$ is the {\it non-vanishing mass current density} associated with the velocity $v^{i}$ of the mass density~\cite{wald84}. 
To be more specific, we investigate the predictions of the linearized gravity when the {\it lowest order effects} of the motion of 
sources are taken into account. In this approximation, we proceed to neglect stresses to yield the energy stress tensor of the form~\cite{wald84}
\beq
T_{\alpha\beta}=2t_{(\alpha}J_{\beta)}-c^{2}\rho_{M}t_{\alpha}t_{\beta},
\label{energystresstensor}
\eeq
where $t^{\alpha}=(\pa/\pa x^{0})^{\alpha}$ is the time direction of the coordinate system and $J_{\beta}=-T_{\alpha\beta}t^{\alpha}$ is mass energy current density four-vector. Note that, in this Wald approximation in the MLGR, we find $J_{\alpha}=\rho_{M}u_{\alpha}$ with $u^{\alpha}=(c, v^{i})$. 
Maintaining the Wald idea associated with (\ref{energystresstensor}), we construct $T^{\alpha\beta}$ in terms of 
$J^{i}_{M}=\rho_{M} v^{i}$ which is linear order in $v^{i}$
\beq
T^{\alpha\beta}=\left(\begin{array}{cc}
c^{2}\rho_{M} &cJ^{j}_{M}\\
cJ^{i}_{M} &0
\end{array}
\right)
\label{energystresstensor2}
\eeq
which reproduces $T^{00}$ and $T^{0i}$ in (\ref{barht0}) and (\ref{barht2}). Note that the space-space component $T^{ij}$ 
in (\ref{energystresstensor2}) vanishes as expected, and the corresponding $\bar{h}^{ij}$ vanishes as in (\ref{hijmlge}) in the Wald approximation. Note also that this energy stress tensor has been used in the Newtonian gravity in (\ref{newtongr}) for the case of $T^{00}=c^{2}\rho_{M}$. In the ensuing section the  Wald approximation also will be exploited in the gravitational wave radiation from rotating binary system where the corresponding 
compact object mass moves with $J^{i}_{M}=\rho_{M} v^{i}$ where $v^{i}$ is given by {\it non-vanishing velocity} $(v^{i}v^{i})^{1/2}=a\omega$ along the tangential direction to the circular orbit of radius $a$ as shown in Figure 1(a). Here $\omega$ is angular frequency of the mass. In this work we will thus make use of the Wald approximation in the MLGR 
in the gravitational wave radiation phenomenology.
 
%-------------------to insert black
%\textcolor{black}{
In the MLGR, inserting (\ref{barht2}) 
into (\ref{squarebarh}) produces
\beq
\square A^{i}_{M}=\frac{4\pi G}{c^{2}} J^{i}_{M}.
\label{naphi}
\eeq
Combining (\ref{naphi0}) and (\ref{naphi}), we find the covariant form of the equations of motion 
\beq
\square A^{\alpha}_{M}=\frac{4\pi G}{c^{2}} J^{\alpha}_{M},
\label{naphi2}
\eeq
where $A_{M}^{\alpha}=\left(\frac{\phi_{M}}{c},A_{M}^{i}\right)$ and the four mass current density is defined by 
$J^{\alpha}_{M}=(c\rho_{M},J^{i}_{M})$. Note that the LGC in (\ref{pabarh0}) is rewritten as 
\beq
\pa_{\alpha}A^{\alpha}_{M}=0.
\label{pabarh02}
\eeq
Exploiting $A^{\alpha}_{M}$ in (\ref{barht0}) and (\ref{barht2}), we find 
\beq
A^{\alpha}_{M}=-\frac{c}{4}\bar{h}^{0\alpha},
\label{aalpham}
\eeq
to yield the equation of motion 
\beq
\square \bar{h}^{0\alpha}=-\frac{16\pi G}{c^{3}} J^{\alpha}_{M}.
\label{nhphi3}
\eeq
Note that the LGC in (\ref{pabarh02}) becomes
\beq
\pa_{\alpha}\bar{h}^{0\alpha}=0.
\label{nhphi4}
\eeq
%}
%-------------------

%%%%%%%%%%%%%%%%%%%%%%%%%%%%%%%%%%%%%%%%%%%%%%%%%%%%%%%%%%%%%%%%%%%%%%%%%%%%%
\begin{figure}[t]
\centering
\vskip -1.0cm
\includegraphics[width=7.0cm]{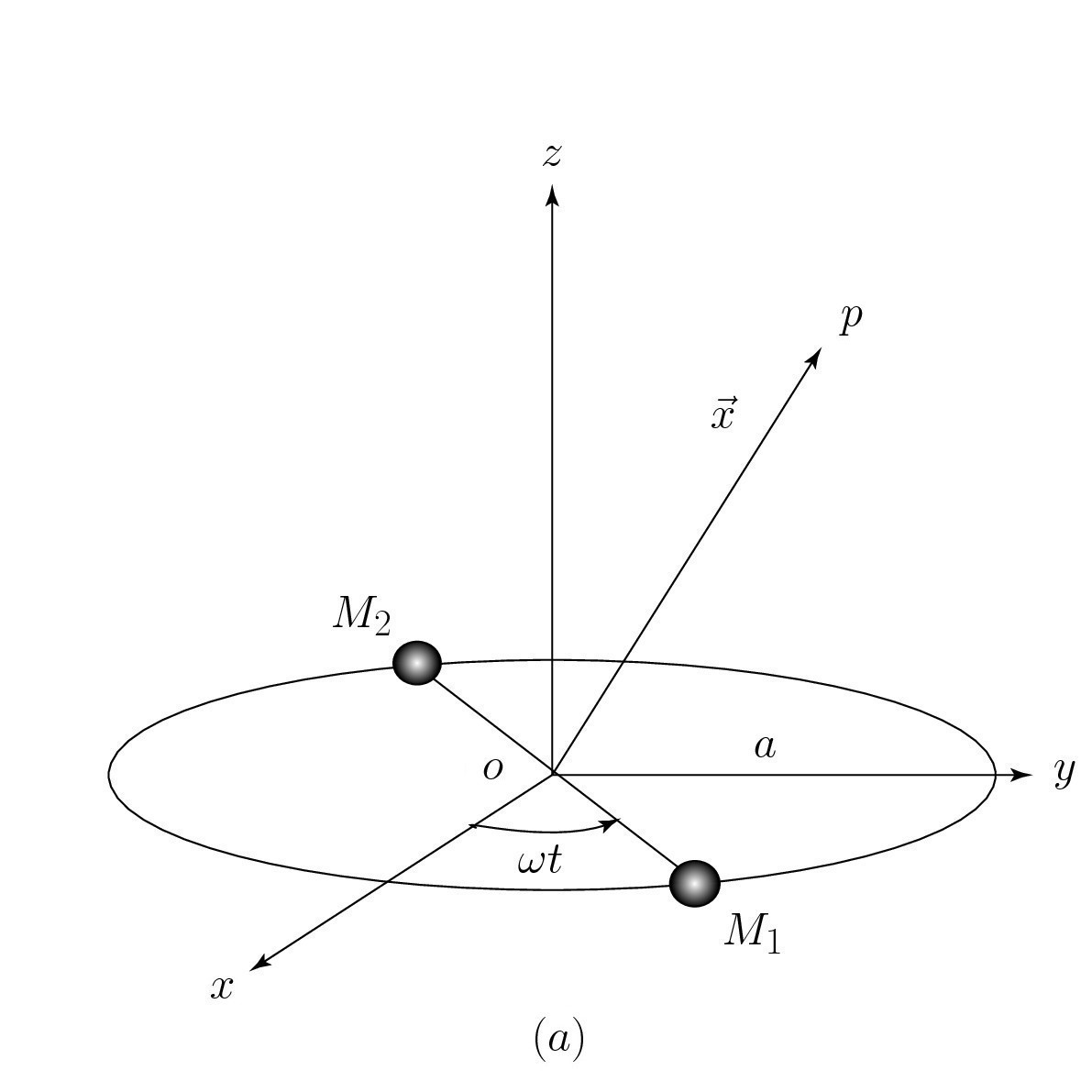}
\hskip 1cm 
\includegraphics[width=7.0cm]{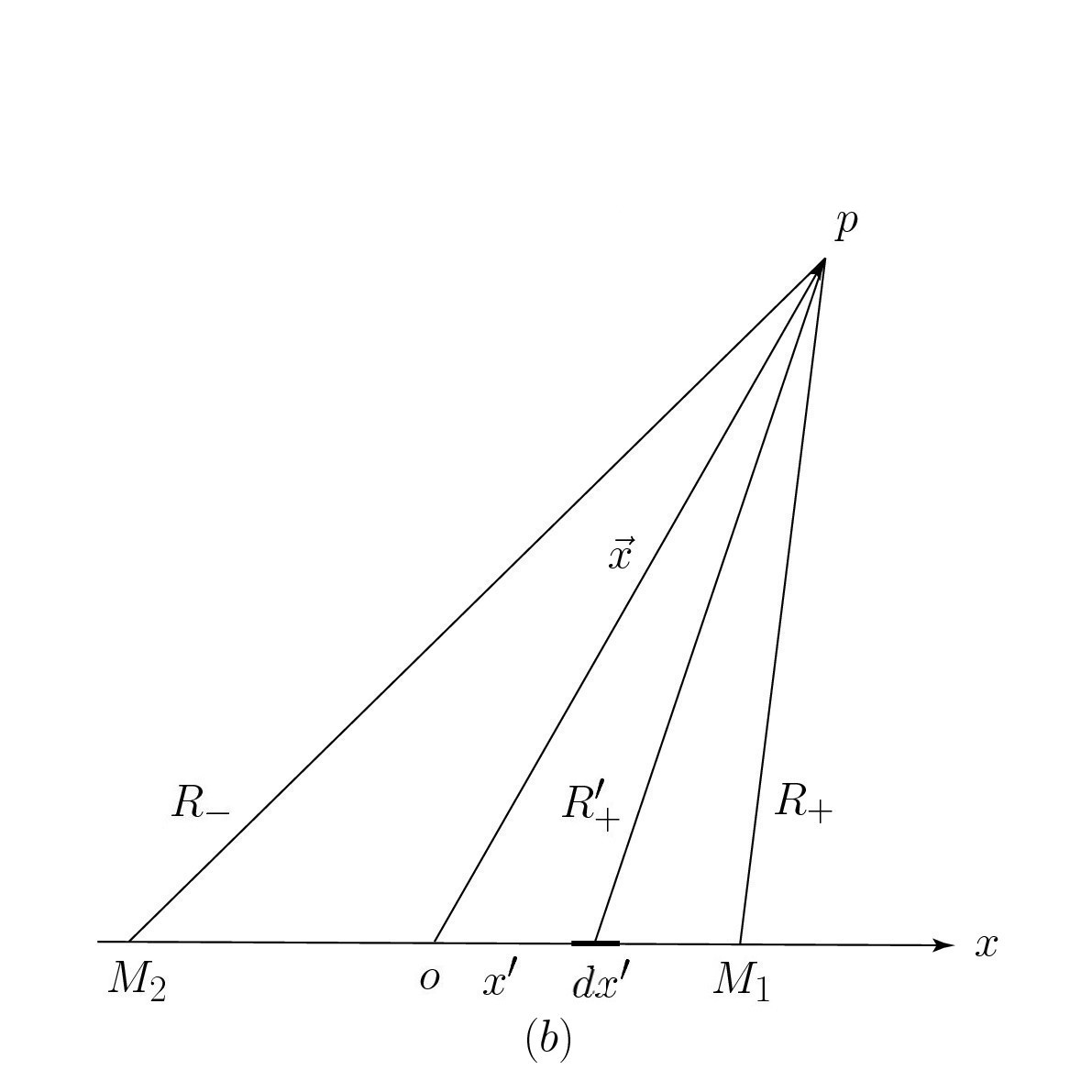}
\vskip -0.6cm
\caption[gw1] {(a) The schematic diagram for a binary system of compact objects $M_{1}$ and $M_{2}$ of equal mass $M$ which 
co-rotate in circular orbits of radius $a$ about their common center of mass $o$ located at the origin, 
with an angular frequency $\omega$. (b) The diagram for the geometry of $R_{\pm}$ for the retarded 
mass scalar potentials $\phi_{M}^{I}$ in (\ref{retpot10}) and $R_{+}^{\prime}$ for the retarded mass 
vector potentials $\vec{A}_{M}^{I}$ in (\ref{am1}). 
Here $\vec{x}=r(\hat{x}\sin\theta\cos\phi+\hat{y}\sin\theta\sin\phi+\hat{z}\cos\theta)$ 
denotes a radial vector from the origin $o$ to an observation point.}
\label{gw1}
\end{figure}
%%%%%%%%%%%%%%%%%%%%%%%%%%%%%%%%%%%%%%%%%%%%%%%%%%%%%%%%%%%%%%%%%%%%%%%%%%%%%

%-------------------to insert 2
%\textcolor{blue}{
Next in the EM phenomenology we obtain the equation of motion 
\beq
\square A_{Q}^{\alpha}=-\frac{1}{c^{2}\epsilon_{0}}J_{Q}^{\alpha},
\label{aqalphajq}
\eeq
where $A_{Q}^{\alpha}=\left(\frac{\phi_{Q}}{c},A_{Q}^{i}\right)$ and $J_{Q}^{\alpha}=(c\rho_{Q},J_{Q}^{i})$. Here the minus sign in 
(\ref{aqalphajq}) denotes that the EM interaction is repulsive and the {\it spin-one} photon mediates repulsive charges. Next 
the plus sign in (\ref{naphi2}) implies that the gravity interaction is attractive. Moreover, 
exploiting the mathematical structures of the equations of motion of $A_{Q}^{\alpha}$ and $A_{M}^{\alpha}$, we find that 
the {\it spin-one} graviton mediates attractive masses. In other words, to treat the spin-one graviton in the 
Wald approximation in the MLGR, we use the vectorial algorithm in (\ref{naphi2}), which is similar to the EM phenomenology discussed 
in (\ref{aqalphajq}). Note that in the LGR having the tensorial 
non-vanishing $\bar{h}_{ij}$ $(i,j=1,2,3)$ there exists no spin-one graviton DOF~\cite{wald84}. To be more specific, in the LGR the gauge transformation is given by 
(\ref{harrowhp}) which has the Lie derivative 
$\pounds_{\xi}\eta_{\alpha\beta}=\pa_{\alpha}\xi_{\beta}+\pa_{\beta}
\xi_{\alpha}$. In contrast, we cannot introduce this kind of second rank tensorial form in the gauge transformation in the MLGR.
%}
%-------------------to insert footnote
\footnote{
%\textcolor{blue}{
In the Wald approximation in the MLGR, there is no way to define the tensorial gauge 
transformation for the non-vanishing $\bar{h}_{0i}$ $(i=1,2,3)$, since the radiation gauge in the 
LGR having the tensorial gauge transformation in (\ref{harrowhp}) already possesses 
the vanishing components $\bar{h}_{0i}=h_{0i}=0$ in (\ref{hohoii}) for instance. To circumvent this difficulty, in 
the Wald approximation we exploit the vectorial gauge transformation in (\ref{barhalphaarr}).}
%}
%-------------------to insert 2
%\textcolor{blue}{
Instead, motivated by the mathematical similarity in the vectorial forms of the equations of motion for 
$A_{M}^{\alpha}\left(=-\frac{c}{4}\bar{h}^{0\alpha}\right)$ $(\alpha=0,1,2,3)$ in (\ref{naphi2}) and $A_{Q}^{\alpha}$ 
in (\ref{aqalphajq}), we make use of the {\it vectorial} gauge transformation 
\beq
\bar{h}_{0\alpha}\rightarrow \bar{h}^{\prime}_{0\alpha}=\bar{h}_{0\alpha}+\pa_{\alpha}\psi,
\label{barhalphaarr}
\eeq
which is {\it analogous} to that in 
the EM phenomenology~\cite{wald84,jackson99} possessing two transverse polarization DOF of the 
spin-one photon and the vectorial gauge transformation
\beq
A_{Q}^{\alpha}\rightarrow A_{Q}^{\alpha\prime}=A_{Q}^{\alpha}+\pa^{\alpha}\zeta.
\label{vectorialq}\eeq
These features imply that, in the Wald approximation in the MLGR, 
the graviton also has spin-one and the corresponding two transverse polarization DOF.
Note that in the LGR we find the spin-two graviton having two transverse polarization DOF in the previous subsection.
%}
%------------------- 

Now we investigate the GW defined in the vacuum. To accomplish this, inserting $J^{\alpha}_{M}=0$ 
into (\ref{nhphi3}) and exploiting $\bar{h}^{0\alpha}$ in (\ref{aalpham}), we obtain the wave equation for 
the non-vanishing field $\bar{h}^{0\alpha}$ (or $A_{M}^{\alpha}$)
\beq
\square \bar{h}^{0\alpha}=\square A_{M}^{\alpha}=0.
\label{squarebarh5}
\eeq
Note that the wave equations in (\ref{squarebarh5}) in the MLGR describe a massless spin-one 
graviton propagating in the flat spacetime, similar to a massless spin-one photon propagating in the flat 
spacetime. In other words, (\ref{squarebarh5}) explains the spin-one gravitons of gravitational field.

Next it seems appropriate to comment on the spin of the graviton in the MLGR. 
In vacuum with $T_{\alpha\beta}=0$ in (\ref{squarebarh}), we end up with the wave equation 
$\square \bar{h}_{\alpha\beta}=0$ which {\it could} describe a massless spin-two 
graviton propagating in the flat spacetime~\cite{pauli39}. However, if we consider only the {\it physically meaningful wave 
equation} for $A^{\alpha}_{M}$ in (\ref{squarebarh5}) obtained from that for $\bar{h}^{0\alpha}$ $(\alpha=0,1,2,3)$, 
we mathematically find a massless spin-one graviton whose properties are well defined in the MLGR. 
Note that the condition $\bar{h}^{ij}=0$ in (\ref{hijmlge}) reduces the number of independent components of 
$\bar{h}^{\alpha\beta}$ to four. The {\it independent} components $(\bar{h}^{00},\bar{h}^{0i})$ thus correspond to the four components of 
$A^{\alpha}_{M}$. Next we explicitly find the solution of the GW field equation in (\ref{squarebarh5}) as follows
\beq
\bar{h}^{0\alpha}=\epsilon^{\alpha}e^{ik_{\mu}x^{\mu}},
\label{solutionh}
\eeq
where $k^{\mu}$ is the GW four-vector. Here $\epsilon^{\alpha}$ is the constant amplitude vector corresponding to the GW 
polarizations.

%-------------------to insert 3
%\textcolor{blue}{  
Now we find that, similar to (\ref{vectorialq}) in the 
EM formalism, the LGC in (\ref{nhphi4}) is preserved by the gauge transformation in (\ref{barhalphaarr}) in the Wald approximation in the 
MLGR if 
\beq
\square \psi=0.
\label{squarepsi}
\eeq
%}
%-------------------
Making use of the relation 
\beq
\pa^{\alpha}\bar{h}_{0\alpha}\rightarrow \pa^{\alpha}\bar{h}^{\prime}_{0\alpha}=\pa^{\alpha}\bar{h}_{0\alpha}
+\square \psi.
\label{lorentz233} 
\eeq
we observe that, to preserve the LGC of $\pa^{\beta}\bar{h}^{\prime}_{\alpha\beta}=0$ in (\ref{lorentz233}) in the MLGR, we need to ensure that 
$\square \psi=0$ as shown in (\ref{squarepsi}). Note that the LGC in (\ref{nhphi4}) in the MLGR has the same structure as the LGC in the electromagnetism.
Note also that the polarization vector $\epsilon^{\alpha}$ has four different components. 
However the one LGC in (\ref{nhphi4}) reduces the number of independent DOF to three. Next the one gauge transformation condition 
in (\ref{squarepsi}) reduces the three independent components to two DOF which implies that we obtain two possible polarizations for the spin-one plane 
GW in (\ref{solutionh}). Note that the gauge choices (\ref{nhphi4}) and (\ref{squarepsi}) in 
the MLGR determine the fundamental nature of the DOF to yield two transverse polarizations, as in the gauge 
choices in (\ref{pabarh0}) and (\ref{squarexi}) in LGR where we also have two transverse polarizations.

Next we study the radiation gauge, and the corresponding initial value problem and dynamics 
of the massless spin-one graviton whose mathematical structure is the same as the massless photon. For the case of massive spin-one graviton, see below (\ref{kalphaneq}). 
Treating the spin-one GW, we exploit the gauge DOF in a source free region having 
$J_{M}^{\alpha}=0$ by using the radiation (or Coulomb) gauge~\cite{wald84,jackson99}. On a constant time surface $t=t_{0}$ of the 
global inertial coordinate system we need to solve
\beq
\pa^{i}\bar{h}_{0i}+\nabla^{2}\psi=0.
\label{delpsidela}
\eeq
We define $\psi$ throughout spacetime to be the solution of (\ref{squarepsi}) whose initial value on the time surface $t=t_{0}$ 
is given by (\ref{delpsidela}). Moreover 
the time derivative of $\psi$ is given by
\beq
\bar{h}_{00}+\frac{\pa \psi}{\pa t}=0.
\label{napsidt}
\eeq
The function defined as 
\beq
F\equiv\bar{h}_{00}+\frac{\pa \psi}{\pa t}
\label{capF}
\eeq
then satisfies the inhomogeneous equation of motion including the gauge DOF
\beq
\square F=-\frac{16\pi G}{c^{3}}J_{M}^{0}.
\label{nablaF}
\eeq
Moreover on the initial surface $t=t_{0}$ we find
\beq
F=0,~~~
\frac{\pa F}{\pa t}=\pa_{i}\bar{h}_{0i}+\nabla^{2}\psi.
\label{labelFanddFdt}
\eeq
Note that, for the inhomogeneous equation of motion having $J_{M}^{\alpha}\neq 0$ $(\alpha=0,1,2,3)$ associated with mass $M$, see the ensuing section where 
$\bar{h}^{0\alpha}(= -\frac{4}{c}A_{M}^{\alpha})$, and the corresponding mass electric and mass magnetic fields $\vec{E}_{M}$ 
and $\vec{B}_{M}$ in (\ref{eb}) will be explicitly evaluated. Moreover the 
directions of $\vec{E}_{M}$ and $\vec{B}_{M}$ will be shown to represent the two transverse directions of GW. In this formulation we do not need the 
remnant components $\bar{h}^{ij}$ in (\ref{hijmlge}) which are unnecessary in the {\it photon-like} spin-one graviton algorithm having the mass scalar and mass 
vector potentials $A_{M}^{\alpha}=(\frac{1}{c}\phi_{M},A_{M}^{i})$. Note also that, for the case of the inhomogeneous equation of motion possessing 
$\rho_{M}\neq 0$ only  which is related with mass $M$ in the TT gauge in the LGR, see Appendix B where we will formulate the quadrupole moment tensor $I^{ij}$ and the 
corresponding second rank tensor $\bar{h}_{TT}^{ij}$ in (\ref{httij2}).

Inserting (\ref{solutionh}) into (\ref{squarebarh5}), we find $k_{\mu}k^{\mu}=0$ to yield
\beq
k^{\mu}=(k,0,0,k),
\label{kalpha000}
\eeq
which describes the massless spin-one graviton. Making use of (\ref{solutionh}) and the LGC in (\ref{nhphi4}), we obtain 
\beq
k_{\mu}\epsilon^{\mu}=0,
\label{kepsilon0}
\eeq 
to produce $\epsilon^{0}=\epsilon^{3}$. Note that the polarization vector $\epsilon^{\alpha}$ satisfies the transversality condition in (\ref{kepsilon0}), 
which is needed since the massless spin-one graviton possesses the transverse components only. Note also that, in the LGR related with
$h_{\alpha\beta}$ in (\ref{planewavegra}) satisfying the LGC in (\ref{pabarh0}), we find the transversality condition $k^{\mu}H_{\mu\nu}=0$ 
in (\ref{kmukmu1}) for the symmetric second rank tensor $H_{\mu\nu}$.

Next we assume $\psi=\epsilon e^{ik_{\mu}x^{\mu}}$ and $\bar{h}^{\prime 0\alpha}=\epsilon^{\prime\alpha}
e^{ik_{\mu}x^{\mu}}$ in the Wald approximation in the MLGR, and then we exploit the gauge transformation equation $\bar{h}^{\prime 0\alpha}=\bar{h}^{0\alpha}+\pa^{\alpha}\psi$ to 
yield 
\beq
\epsilon^{\prime\alpha}=\epsilon^{\alpha}+i\epsilon k^{\alpha}.
\label{epsilon123}
\eeq
Using (\ref{kalpha000}) and (\ref{epsilon123}), we explicitly find the non-trivial relation 
$\epsilon^{\prime 0}=\epsilon^{0}+i\epsilon k$. Next, exploiting the ansatz $\epsilon=\frac{i}{k}\epsilon^{0}$, we find 
$\epsilon^{\prime 0}=0$. Dropping the primes in $\epsilon^{\prime\alpha}$ we obtain 
\beq
\epsilon^{\alpha}=(0,\epsilon^{1},\epsilon^{2},0),
\label{epsilon0120}
\eeq
implying that the GW is transverse to the propagation direction $k^{\alpha}$ in (\ref{kalpha000}). Note that 
$\epsilon^{\alpha}$ in (\ref{epsilon0120}) thus satisfies the identity $\epsilon^{\alpha}k_{\alpha}=0$ in (\ref{kepsilon0}) 
as expected.

Now we have comments on the GW phenomenology in terms of $\epsilon^{\alpha}$ in (\ref{epsilon0120}) in the Wald approximation in the MLGR. First, in the cases of $\epsilon^{\alpha}=(0,\epsilon^{1},0,0)$ and $\epsilon^{\alpha}=(0,0,\epsilon^{2},0)$, 
as the GW passes a free test particle, the test particle will oscillate in the 
$x$ and $y$ directions with a magnitude changing sinusoidally with time, respectively, following the GW wave relation 
in (\ref{solutionh}). Second, in the cases of $\epsilon^{\alpha}=(0,\epsilon^{1},i\epsilon^{1},0)$ and 
$\epsilon^{\alpha}=(0,\epsilon^{1},-i\epsilon^{1},0)$, the GW polarizations yield left-circularly and right-circularly polarized GWs, 
and then the test particle will move in the corresponding circles, respectively. 
Third, in the general cases of $\epsilon^{\alpha}=(0,\epsilon^{1},i\bar{\epsilon}^{1},0)$ and 
$\epsilon^{\alpha}=(0,\epsilon^{1},-i\bar{\epsilon}^{1},0)$ $(\epsilon^{1}\neq \bar{\epsilon}^{1})$, the GW polarizations produce 
left-elliptic and right-elliptic polarized GWs, respectively, and thus the test particle will move in the corresponding ellipses.

Next we investigate the properties of the massive graviton in the Wald approximation in the MLGR. As in the case of the massive photon~\cite{hong22qm}, we find the identity 
\beq
k_{\mu}\epsilon^{\mu}\neq 0,
\label{kalphaneq}
\eeq
for the massive graviton, since for the massive graviton we have longitudinal component in addition to transverse ones, 
similar to the phonon associated with massive particle lattice vibrations~\cite{phonon}. 
Namely, the massive spin-one graviton has three components to produce three DOF, similar to the massive photon and to the massive spin-two graviton. 
Note that the dynamics of the massive spin-one graviton are similar to those of the spin-two graviton. 
Note also that  $\epsilon^{\alpha}$ is a polarization vector 
possessing the spacetime index $\alpha$ $(\alpha=0,1,2,3)$ which is the same as the massive graviton spin index and 
is needed to incorporate minimally the spin DOF for the massive graviton. Next, as 
in the case of the GRB in the Dirac type relativistic massive photon model~\cite{hong22qm}, the massive graviton could 
be interpreted as a spin-one GRB-like particle~\cite{grb1,goldstein2017}. The GRB-like graviton 
corresponding to its negative energy solution could then be regarded as an anti-graviton, which could be 
a candidate for an intense radiation flare of the GRB of GRB170817A~\cite{grb1,goldstein2017} associated with the binary compact objects merger event GW170817. 
Note that the massive anti-graviton could be described in terms of 
the massive gauge boson possessing a finite radius as in the massive photon in the stringy photon model~\cite{hong211}. 
For more details of the formalism of massive spin-one graviton, see Appendix C.

%%%%%%%%%%%%%%%%%%%%%%%%%%%%%%%%%%%%%%%%%%%%%%%%%%%%%%%%%%%%%%%%%%%%%%%%%%
\section{Gravitational wave radiation from merging binary compact objects}
%setcounter{equation}{0}
\renewcommand{\theequation}{\arabic{section}.\arabic{equation}}
\label{setupgravitosection}
%%%%%%%%%%%%%%%%%%%%%%%%%%%%%%%%%%%%%%%%%%%%%%%%%%%%%%%%%%%%%%%%%%%%%%%%%%

In this section, in the MLGR we will investigate a binary system of merging compact objects which possesses masses $M_{1}$ 
and $M_{2}$ rotating  on the $x$-$y$ plane. We assume that for simplicity the two compact objects $M_{1}$ and $M_{2}$ 
of equal mass $M_{1}=M_{2}=M$ co-rotate in circular orbits of radius $a$ 
about their common center of mass $o$ with an angular frequency $\omega$, as shown in Figure 1(a).
Note that the mass moves with {\it non-vanishing velocity} $(v^{i}v^{i})^{1/2}=a\omega$ along the tangential 
direction to the circular orbit as shown in Figure 1(a) which is consistent with the Wald approximation in the MLGR discussed in the previous section.
Note also that the rotating binary compact objects can be thought of as the superposition of two oscillating binary compact objects, 
one along the $x$-axis and the other along the $y$-axis. The rotating mass vector $\vec{M}(t)$ on the $x$-$y$ plane is given by
\beq
\vec{M}(t)=M(\hat{x}\cos\omega t+\hat{y}\sin\omega t).
\label{atvector}
\eeq
For the masses on the $x$-axis and $y$-axis, exploiting (\ref{atvector}) the Green retarded~\cite{wald84,hobson06} 
mass scalar potentials (or gravitational potentials) $\phi_{M}^{I}$ and $\phi_{M}^{II}$ are respectively given by
\bea
\phi_{M}^{I}(\vec{x},t)&=&-G\left(\frac{M\cos[\omega(t-R_{+}/c)]}{R_{+}}+\frac{M\cos[\omega(t-R_{-}/c)]}{R_{-}}\right),\nn\\
\phi_{M}^{II}(\vec{x},t)&=&-G\left(\frac{M\sin[\omega(t-\bar{R}_{+}/c)]}{\bar{R}_{+}}
+\frac{M\sin[\omega(t-\bar{R}_{-}/c)]}{\bar{R}_{-}}\right),
\label{retpot10}
\eea
where 
\bea
R_{\pm}&=&r\left(1\mp \frac{2a}{r}\sin\theta\cos\phi+\frac{a^{2}}{r^{2}}\right)^{1/2}
= r\left(1\mp \frac{a}{r}\sin\theta\cos\phi\right),\nn\\
\bar{R}_{\pm}&=&r\left(1\mp \frac{2a}{r}\sin\theta\sin\phi+\frac{a^{2}}{r^{2}}\right)^{1/2}
=r\left(1\mp \frac{a}{r}\sin\theta\sin\phi\right).
\label{rpm}
\eea
Here we have used approximation $a\ll r$. For more details of $R_{\pm}$, see the geometry in Figure 1(b). 
After some algebra we arrive at the total contributions to the mass scalar potentials 
$\phi_{M}=\phi_{M}^{I}+\phi_{M}^{II}$
\bea
\phi_{M}(\vec{x},t)&=&-\frac{2GM}{r}\left(\cos\left[\omega\left(t-\frac{r}{c}\right)\right]
+\sin\left[\omega\left(t-\frac{r}{c}\right)\right]
-\frac{\omega a^{2}}{cr}\sin^{2}\theta\cos^{2}\phi\sin\left[\omega\left(t-\frac{r}{c}\right)\right]\right.\nn\\
&&\left.+\frac{\omega a^{2}}{cr}\sin^{2}\theta\sin^{2}\phi\cos\left[\omega\left(t-\frac{r}{c}\right)\right]\right),
\label{retpottotal}
\eea 
where we have ignored the much smaller terms exploiting approximation $\frac{1}{r}\ll\frac{\omega}{c}$. Note that 
the wavelength $\lambda\left(=\frac{2\pi c}{\omega}\right)$ of the GW  
is also much smaller than $r$ to yield $\lambda\ll r$.

Next we construct the mass vector potential $\vec{A}_{M}$ which is related with the mass current and is missing in the 
Newtonian gravity. Making use of (\ref{atvector}) 
for the binary compact objects co-rotating on the $x$-$y$ plane, we find the mass current 
\beq
\vec{I}_{M}(t)=-M\omega(\hat{x}\sin \omega t-\hat{y}\cos\omega t).
\label{imtcurr}
\eeq 
We then explicitly obtain the retarded mass vector potentials $\vec{A}_{M}^{I}$ and $\vec{A}_{M}^{II}$ 
for the mass current on the $x$-axis and $y$-axis
\bea
\vec{A}_{M}^{I}(\vec{x},t)&=&-\frac{GM\omega}{c^{2}}
\left(\int_{-a}^{0}\frac{\sin[\omega(t-R_{-}^{\prime}/c)]}{R_{-}^{\prime}}dx^{\prime}
-\int_{0}^{\bar{a}}\frac{\sin[\omega(t-R_{+}^{\prime}/c)]}{R_{+}^{\prime}}dx^{\prime}\right)\hat{x},\nn\\
\vec{A}_{M}^{II}(\vec{x},t)&=&-\frac{GM\omega}{c^{2}}
\left(-\int_{-a}^{0}\frac{\cos[\omega(t-\bar{R}_{-}^{\prime}/c)]}{\bar{R}_{-}^{\prime}}dy^{\prime}
+\int_{0}^{\bar{a}}\frac{\cos[\omega(t-\bar{R}_{+}^{\prime}/c)]}{\bar{R}_{+}^{\prime}}dy^{\prime}\right)\hat{y},
\label{am1}
\eea
where $R_{\pm}^{\prime}$ and $\bar{R}_{\pm}^{\prime}$ are given by
\beq
R_{\pm}^{\prime}=r\left(1\mp \frac{x^{\prime}}{r}\sin\theta\cos\phi\right),~~~
\bar{R}_{\pm}^{\prime}= r\left(1\mp \frac{y^{\prime}}{r}\sin\theta\sin\phi\right).
\label{rpmprime22}
\eeq
Here we also have exploited approximation $a\ll r$. For more details of $R_{+}^{\prime}$, see the geometry in Figure 1(b). 
Note that $\bar{a}$ in the upper bound of the second integral in (\ref{am1}) is the reduced radius from the origin after the mass 
$M_{1}(=M)$ travels during the half revolution. The distance $\bar{a}$ is defined by 
\beq
\delta a\equiv 2(a-\bar{a}),
\label{deltaadefn}
\eeq
so that $\delta a$ can measure the distortion of the spirally merging binary compact objects during one cycle of their revolution.
Note also that in the GW radiation phenomenology we assume that $\delta a/a$ is extremely small to produce almost 
no stress in the two rotating mass compact objects. In the two body system of the rotating compact objects, 
the mass velocity ${(v^{i}v^{i})}^{1/2}$ $(=a\omega)$ is meaningful to implies that the Wald approximation in the MLGR defined in 
(\ref{energystresstensor}) and 
(\ref{energystresstensor2}) is well proposed.

After some algebra associated with the definite integrations, we arrive at the total retarded mass vector potential
$\vec{A}_{M}=\vec{A}_{M}^{I}+\vec{A}_{M}^{II}$
\bea
\vec{A}_{M}(\vec{x},t)&=&-\frac{GM\omega\delta a}{2c^{2}r}\left[\left(
\sin\left[\omega\left(t-\frac{r}{c}\right)\right]+\frac{\omega a}{c}\sin\theta\cos\phi
\cos\left[\omega\left(t-\frac{r}{c}\right)\right]\right)\hat{x}\right.\nn\\
&&\left.+\left(
-\cos\left[\omega\left(t-\frac{r}{c}\right)\right]+\frac{\omega a}{c}\sin\theta\sin\phi
\sin\left[\omega\left(t-\frac{r}{c}\right)\right]\right)\hat{y}\right],
\label{amtotal}
\eea 
where we have again used the approximations $a \ll r$ and $\lambda\ll r$ with $\lambda$ being the GW wavelength. 
Note that, for the binary system of the {\it two masses $M_{1}=M_{2}=M$ of interest},\footnote{This binary system phenomenology 
can be applicable to an isolated binary system of the {\it particle} masses $M_{1}=M_{2}=M$.}
we obtain $\bar{h}^{\alpha\beta}$ of the form
\beq
\bar{h}^{00}=-\frac{4}{c^{2}}\phi_{M}(\vec{x},t),~~~\bar{h}^{0i}=-\frac{4}{c}A_{M}^{i}(\vec{x},t),~~~\bar{h}^{ij}=0,
\label{hbarbinary}
\eeq
where we have used (\ref{hijmlge}), (\ref{h00phim}) and (\ref{barht2}). Here $\phi_{M}(\vec{x},t)$ and $A_{M}^{i}(\vec{x},t)$ are 
given by (\ref{retpottotal}) and (\ref{amtotal}), respectively. Using the relation 
$h_{\alpha\beta}=\bar{h}_{\alpha\beta}-\frac{1}{2}\eta_{\alpha\beta}\bar{h}$ where 
$\bar{h}=\eta^{\alpha\beta}\bar{h}_{\alpha\beta}$, we can define the specific gravitational spacetime 4$\times$4 metric tensor 
$h_{\alpha\beta}$ $(\alpha, \beta=0,1,2,3)$ of the mass binary system.

Now exploiting (\ref{retpottotal}) and (\ref{amtotal}), we find the mass electric field which also includes 
the mass vector potential effect $\vec{E}_{M}\equiv  -\na\phi_{M}-\frac{\pa \vec{A}_{M}}{\pa t}=E_{M}^{r}\hat{r}+E_{M}^{\theta}\hat{\theta}
+E_{M}^{\phi}\hat{\phi}$ and the mass magnetic one $\vec{B}_{M}\equiv \na\times\vec{A}_{M}=B_{M}^{\theta}\hat{\theta}+B_{M}^{\phi}\hat{\phi}$
\bea
E_{M}^{r}&=&\frac{2GM\omega}{cr}
\left(\sin\left[\omega\left(t-\frac{r}{c}\right)\right]-\cos\left[\omega\left(t-\frac{r}{c}\right)\right]\right),\nn\\
E_{M}^{\theta}&=&cB_{M}^{\phi}=\frac{GM\omega^{2}\delta a}{2c^{2}r}\cos\theta\left(\cos\left[\omega\left(t-\frac{r}{c}\right)\right]\cos\phi
+\sin\left[\omega\left(t-\frac{r}{c}\right)\right]\sin\phi\right),\nn\\
E_{M}^{\phi}&=&-cB_{M}^{\theta}=\frac{GM\omega^{2}\delta a}{2c^{2}r}\left(-\cos\left[\omega\left(t-\frac{r}{c}\right)\right]\sin\phi
+\sin\left[\omega\left(t-\frac{r}{c}\right)\right]\cos\phi\right),
\label{eb}
\eea 
where we have included the relevant leading order terms. Exploiting (\ref{eb}), we find $\vec{B}_{M}=\frac{1}{c}(\hat{r}\times\vec{E}_{M})$ 
which implies that the $\vec{E}_{M}$ and $\vec{B}_{M}$ fields are mutually perpendicular, and are spherical waves (not plane waves) and 
their amplitudes decrease like $1/r$ as they propagate. However, for large $r$, the GW waves are approximately plane waves over 
small regions, similar to the EM waves in (\ref{qeb}) discussed in Appendix A. Note that the $\vec{E}_{M}$ field possesses 
$E_{M}^{r}\hat{r}$ component originated from the gravitational attraction from the binary compact objects. This characteristic is 
different from the rotating charge electric dipole moment system where $E_{Q}^{r}\hat{r}$ component vanishes as in (\ref{qeb}).

Next we formulate the mass Poyinting vector in terms of $\vec{E}_{M}$ and $\vec{B}_{M}$ in (\ref{eb}) 
\beq
\vec{S}_{M}\equiv\frac{c^{2}}{4\pi G}(\vec{E}_{M}\times\vec{B}_{M})
=\frac{c}{4\pi G}\left[(E_{M}^{\perp})^{2}\hat{r}-E_{M}^{r}\vec{E}_{M}^{\perp}\right],
\label{poynting}
\eeq
where 
\bea
\vec{E}_{M}^{\perp}&=&E_{M}^{\theta}\hat{\theta}+E_{M}^{\phi}\hat{\phi},\nn\\
(E_{M}^{\perp})^{2}&=&(E_{M}^{\theta})^{2}+(E_{M}^{\phi})^{2}=\left(\frac{GM\omega^{2}\delta a}{2c^{2}r}\right)^{2}\left(1-\sin^{2}\theta\cos^{2}
\left[\omega\left(t-\frac{r}{c}\right)-\phi\right]\right).
\label{poynting2}
\eea
Note that the normalization factor $\frac{c^{2}}{4\pi G}$ in (\ref{poynting}) is systematically determined by using the 
proportionality constants $-\frac{G}{c^{2}}$ of $\vec{A}_{M}^{I}$ and $\vec{A}_{M}^{II}$ in (\ref{am1}), 
and $-G$ of $\phi_{M}^{I}$ and $\phi_{M}^{II}$  in (\ref{retpot10}) which is consistent with the Newtonian 
gravity~\cite{goldstein80,wald84}. Note also that, in the optics associated with the massless spin-one graviton, the mass electric and mass magnetic fields 
$\vec{E}_{M}$ and $\vec{B}_{M}$ represent the two transverse directions of the GW, and the mass Poyinting vector $\vec{S}_{M}$ is parallel to the direction of the GW.

Using (\ref{poynting}) and (\ref{poynting2}) we find the GW radiation intensity, namely the GW radiation power per surface $\Sigma$, 
which is obtainable by averaging in time over a complete cycle
\beq
\frac{dP_{M}}{d\Sigma}\equiv\langle\vec{S}_{M}\rangle=\frac{c}{4\pi G}
\left(\frac{GM \omega^{2}\delta a}{2c^{2}r}\right)^{2}\left(1-\frac{1}{2}\sin^{2}\theta\right)\hat{r}.
\label{poynting3}
\eeq
Note that in (\ref{poynting3}) the $\vec{E}_{M}^{\perp}$-component of $\vec{S}_{M}$ in (\ref{poynting}) vanishes, after 
averaging in time over a complete cycle and then using $\langle E_{M}^{r}\rangle=0$. 
To be specific, in the MLGR we find the non-vanishing $\phi_{M}(=-\frac{c^{2}}{4}\bar{h}^{00})$ in (\ref{retpottotal}) and the corresponding $E_{M}^{r}\hat{r}=-\nabla\phi_{M}$ in (\ref{eb}). However, after averaging in time over a complete cycle, we have the vanishing contribution of $\bar{h}^{00}$ to the {\it physical quantity} $\langle E_{M}^{r}\rangle$ in the GW radiation algorithm. 
%-------------------to insert 4
%\textcolor{blue}{
In this {\it phenomenological} sense, this aspect of $\bar{h}^{00}$ in the Wald approximation in the MLGR is similar to that of 
$\bar{h}_{TT}^{00}=0$ 
in (\ref{standardlgr}) in the TT gauge in the LGR, even though the gauge transformation in (\ref{barhalphaarr}) 
in the MLGR vectorial algorithm is different from that in (\ref{harrowhp}) in the LGR tensorial scheme. 
Moreover, we have another {\it phenomenological} characteristic that the two transverse directions of the GW are perpendicular to the 
GW radiation direction in the Wald approximation, similar to those  in the LGR.
%}
%-------------------
Note also that, we have the vanishing contribution of $E_{M}^{r}\hat{r}$ to 
the GW radiation intensity $\frac{dP_{M}}{d\Sigma}$ in (\ref{poynting3}) 
by averaging in time over a complete cycle.

%%%%%%%%%%%%%%%%%%%%%%%%%%%%%%%%%%%%%%%%%%%%%%%%%%%%%%%%%%%%%%%%%%%%%%%%%%%%%
\begin{figure}[t]
\centering
\vskip -1.0cm
\includegraphics[width=7.0cm]{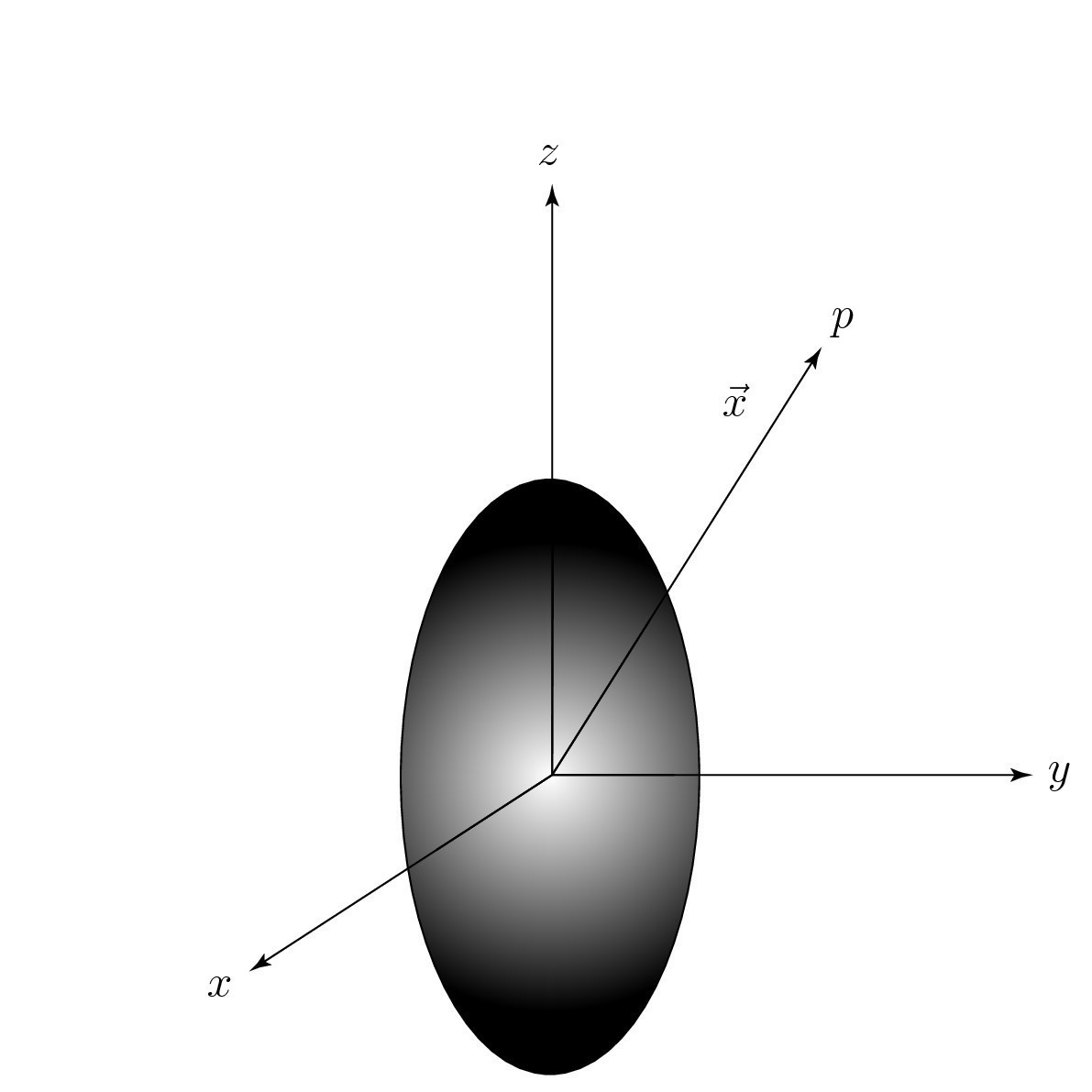}
%\vskip -0.6cm
\vskip -0.3cm
\caption[gw2] {The characteristic diagram for a surface of the GW radiation intensity $\frac{dP_{M}}{d\Sigma}$ in (\ref{poynting3}), 
as a function of the polar angle $\theta$, showing a 
prolate ellipsoid geometry in the corresponding intensity profile. Here the binary compact objects reside on the equatorial $x$-$y$ plane as 
shown in Figure 1(a). Note that, at a given radial distance $r=|\vec{x}|$ from the origin of the binary system of compact objects, 
the GW radiation intensity on the revolution $z$ axis of the binary compact objects is twice that on the equatorial plane.}
\label{gw2}
\end{figure}
%%%%%%%%%%%%%%%%%%%%%%%%%%%%%%%%%%%%%%%%%%%%%%%%%%%%%%%%%%%%%%%%%%%%%%%%%%%%%

Now $\frac{dP_{M}}{d\Sigma}$ has a maximum 
(minimum) value at $\theta=0$ ($\theta=\frac{\pi}{2}$) to yield the invariant relation independent of the radial distance $r$
\beq
\frac{dP_{M}}{d\Sigma}(\theta=0)=2\frac{dP_{M}}{d\Sigma}\left(\theta=\frac{\pi}{2}\right),
\label{invidentity}
\eeq
implying that, at a given radial distance $r$ from the origin of the binary system of compact objects, 
the GW radiation intensity on the revolution $z$ axis of the compact objects is {\it twice} that on 
the equatorial $x$-$y$ plane where the compact objects locate. 
The characteristic invariant in (\ref{invidentity}) also occurs in the EM radiation 
intensity of the rotating charge electric dipole moment in (\ref{qpoynting3}). 
The geometrically invariant surface of the GW radiation intensity $\frac{dP_{M}}{d\Sigma}$ 
then yields a prolate ellipsoid geometry in the corresponding intensity profile as shown in Figure 2.

The total GW radiation power constructed by integrating $\langle\vec{S}_{M}\rangle$ over a sphere 
$\Sigma_{r}$ of radius $r$ is then given by
\beq
P_{M}\equiv\int\langle\vec{S}_{M}\rangle\cdot d\vec{\sigma}=\frac{GM^{2}\omega^{4}(\delta a)^{2}}{6c^{3}},
\label{poynting4}
\eeq
where $d\vec{\sigma}$ is an area element vector perpendicular to the surface $\Sigma_{r}$. 
As expected, the dimensionality of $P_{M}$ is that of power, which is the same as the dimensionality 
of $P_{Q}$ in (\ref{qpoynting4}). This aspect implies that the normalization factor $\frac{c^{2}}{4\pi G}$ in (\ref{poynting}) 
is well defined. Note that in the {\it ideal} binary compact objects without any distortion ($\delta a=0$) 
we obtain the vanishing total GW radiation power. In contrast, in the {\it physically merging} binary compact objects source we have 
the total power in (\ref{poynting4}) which is proportional to $(\delta a)^{2}$. Here $\delta a$ is defined in (\ref{deltaadefn}). 
These phenomenological aspects of the distortion are consistent with the TT gauge approach to the merging binary compact 
objects~\cite{wald84,hobson06} in the LGR.

Now, it seems appropriate to comment on the TT gauge approach in the LGR, which is distinct from 
the MLGR defined in (\ref{barht0}) and (\ref{barht2}). First, for the LGR associated with merging binary compact objects of 
equal mass $M$ which co-rotate in circular orbits of radius $a$ about their common center of mass $o$ with 
an angular frequency $\omega$ as shown in Figure 1(a), we construct the symmetric second rank tensor for the GW radiation in the TT gauge~\cite{hobson06}
\bea 
\bar{h}_{TT}^{00}&=&0,~~~\bar{h}_{TT}^{0i}=0,\nn\\
\bar{h}^{ij}_{TT}&=&\frac{8GM \omega^{2}a^{2}}{c^{4}r}
\left(
\begin{array}{ccc}
\cos [2\omega(t-r/c)] &\sin [2\omega(t-r/c)] &0\\
\sin [2\omega(t-r/c)] &-\cos [2\omega(t-r/c)] &0\\
0 &0 &0
\end{array}
\right),~~~(i,j=1,2,3).
\label{standardlgr}
\eea
Note that the GW field $\bar{h}_{TT}^{ij}$ in (\ref{standardlgr}) represents spherical wave (not plane wave) and 
its amplitude decreases like $1/r$ as it propagates, as in the case of $\vec{E}_{M}$ and $\vec{B}_{M}$ in (\ref{eb}) in the 
MLGR. 
In order to investigate the GW from binary compact objects in the LGR, we introduce the 
TT gauge~\cite{wald84,hobson06}, and thus the GW originates from the second rank tensor $\bar{h}_{TT}^{\alpha\beta}$ 
in (\ref{standardlgr}). Using $\bar{h}_{TT}^{\alpha\beta}$ in (\ref{standardlgr}), in the TT gauge we find
another condition  
\beq
\bar{h}_{TT}=0,
\label{anotherh}
\eeq
which yields 
\beq
\bar{h}_{TT}^{\alpha\beta}=h_{TT}^{\alpha\beta}.
\label{anotherh2}
\eeq
Exploiting (\ref{standardlgr}) and (\ref{anotherh2}), we can thus construct the spacetime metric 
$g_{TT}^{\alpha\beta}=\eta^{\alpha\beta}+h_{TT}^{\alpha\beta}$ for (\ref{standardlgr}).

Second, the GW radiation from the binary compact objects in the MLGR can be described in 
terms of the mathematically and physically independent vector $\bar{h}^{0\alpha}=-\frac{4}{c}A_{M}^{\alpha}$ $(\alpha=0,1,2,3)$
in (\ref{barht0}) and (\ref{barht2}), without resorting to the second rank tensor $\bar{h}_{TT}^{ij}$ 
in (\ref{standardlgr}) in the LGR. Note also that the graviton possesses spin-one in the MLGR, 
but spin-two in the LGR, since the MLGR and LGR are delineated in terms of the vector $\bar{h}^{0\alpha}$ 
related with $\bar{h}^{ij}=0$ in (\ref{hijmlge}), and the second rank tensor $\bar{h}_{TT}^{\alpha\beta}$ (especially 
the non-vanishing space-space components $\bar{h}_{TT}^{ij}$ of the tensor in (\ref{standardlgr})), respectively. 

Third, in the LGR the second rank tensor $\bar{h}_{TT}^{ij}$ does not include the angle $(\theta,\phi)$ dependences of 
$A_{M}^{i}=-\frac{c}{4}\bar{h}^{0i}$ in (\ref{amtotal}) in the MLGR, which are related with those of 
the GW radiation intensity profile having a prolate ellipsoid geometry in Figure~\ref{gw2}. 
To be more specific, formulating the components $\bar{h}_{TT}^{ij}$ in (\ref{standardlgr}) we have used the $1/r$ factor and the retarded time 
$t^{\prime}=t-r/c$ without considering the $(\theta,\phi)$ dependences of the vector $\vec{x}$ shown in Figute 1(a). In contrast, 
in the Wald approximation in the MLGR, the angle dependences appear explicitly in the distances between the observation point $p$ and 
the masses $M_{1}$ and $M_{2}$ and also in the corresponding retarded time factors in 
$\bar{h}^{00}=-\frac{4}{c^{2}}(\Phi_{M}^{I}+\Phi_{M}^{II})$, as shown in  (\ref{retpot10}) and (\ref{rpm}) for instance.

%-------------------to insert 5
%\textcolor{blue}{
Fourth, in the TT gauge, the DOF of $\bar{h}_{TT}^{\alpha\beta}$ in (\ref{standardlgr}) is two which is consistent with the corresponding DOF discussed in Section 2. Namely this DOF corresponds to two possible polarizations for plane GW associated with the massless spin-two graviton. 
Note that, for the massless plane GW traveling in 
the $z$ direction, we find the GW four-vector $k^{\mu}=(k,0,0,k)$ in (\ref{kkk00}). 
The traceless tensor $\bar{h}_{TT}^{\alpha\beta}$ in (\ref{standardlgr}) in the LGR then 
satisfies the transversality condition $k_{\mu}\bar{h}_{TT}^{\mu\nu}=0$. Note also that, in the Wald approximation in the MLGR, the polarization vector $\epsilon^{\alpha}=(0,\epsilon^{1},\epsilon^{2},0)$ in (\ref{epsilon0120}) and the GW four-vector 
$k^{\mu}=(k,0,0,k)$ in (\ref{kalpha000}) satisfy the 
transversality condition $k_{\mu}\epsilon^{\mu}=0$ in (\ref{kepsilon0}). 
These aspects imply that the {\it phenomenology} of the LGR is similar to that in the Wald approximation in the sense that the two algorithms produce the transversality conditions 
(\ref{kmukmu1}) and (\ref{kepsilon0}) respectively, even though these algorithms are mathematically different to each other. 
For more details of the TT gauge, see Appendix B.
%}
%-------------------

%%%%%%%%%%%%%%%%%%%%%%%%%%%%%%%%%%%%%%%%%%%%%%%%%%%%%%%%%%%%%%%%%%%%%%%%
\section{Conclusions}
\setcounter{equation}{0}
\renewcommand{\theequation}{\arabic{section}.\arabic{equation}}
%%%%%%%%%%%%%%%%%%%%%%%%%%%%%%%%%%%%%%%%%%%%%%%%%%%%%%%%%%%%%%%%%%%%%%%%

In summary, we have formulated the MLGR where the mass scalar potential has the additional dynamical DOF. 
In the static gravitational field limit of the MLGR, the mass scalar potential is consistent with 
the well-established gravitational potential in the Newtonian gravity. Next we also have found that the MLGR allows the mass vector 
potential which originates from the mass current and is missing in the Newtonian gravity. 
As an application of the MLGR, we have investigated the phenomenological aspects of the 
merging binary compact objects which co-rotate in circular orbits about their common center of mass with a constant angular frequency. 
To be specific, we have constructed the mass scalar and mass vector potentials of the merging binary compact objects at an observation 
point located far away from the compact objects. We then have formulated the mass electric and mass magnetic fields, and the mass Poyinting vector. 
Next we have found the GW radiation intensity profile having a prolate ellipsoid geometry due to the merging 
binary compact objects source, and then explicitly obtained the total GW radiation power due to the source. 
We also have observed that, in no distorting limit of the binary compact objects, they do not yield the total GW radiation power, 
which is consistent with the TT gauge approach in the LGR. Note that the MLGR related with 
$\bar{h}^{00}=-\frac{4}{c^{2}}\phi_{M}$ is consistently reduced to the Newtonian gravity having $\na^{2} \phi_{M}=4\pi G\rho_{M}$ in the static limit, 
differently from the TT gauge approach with $\bar{h}^{00}_{TT}=0$. This fact implies that the gauge in the MLGR is applicable to both the GW radiation 
and Newtonian gravity simultaneously. Recently the galaxy-based observatories have been proposed~\cite{dan23} and thus the 
angle $(\theta,\phi)$ dependence of the GW radiation depicted in Figure 1(a), which is not explained in the TT gauge, could become important 
increasingly in the astrophysical phenomenology. To be specific, the MLGR constructed in this work could be exploited in the future experimental 
instruments, which could detect the angle dependence of 
the GW radiation power from the merging binary compact objects. The MLGR algorithm for the 
angle dependence of the GW could then play a role in the realistic {\it precision astrophysics}. 
One of the main points of this paper is that, in the MLGR we have 
found the mathematically and physically well defined spin-one graviton, 
distinct from the second rank tensor algorithm in the LGR which could allow the spin-two graviton. It will be interesting to search 
for the spin-one graviton which could be related with the well-established spin-one photon phenomenology 
such as the photoelectric effect and Compton scattering for instance. Once this is done, the MLGR algorithm associated with 
the graviton could give some progress impacts on the precision astrophysics. 
Assuming the possibility of the {\it massive} spin-one graviton, we have observed that the massive graviton would be interpreted as a 
GRB-like graviton, which could then be regarded as an anti-graviton. This anti-graviton could be phenomenologically 
a candidate for an {\it intense radiation flare} of the GRB170817A related with the {\it binary compact objects merger event} GW170817. 
Finally we have shown that the spin-one graviton in the MLGR has two polarization DOF whose aspect is shared by the 
spin-two graviton in the LGR.

%%%%%%%%%%%%%%%%%%%%%%%%%%%%%%%%%%%%%%%%%%%%%%%%%%%%%%%%%%%%%%%%%%%%%%%%%%%%%
\begin{figure}[t]
\centering
\vskip -1.0cm
\includegraphics[width=7.0cm]{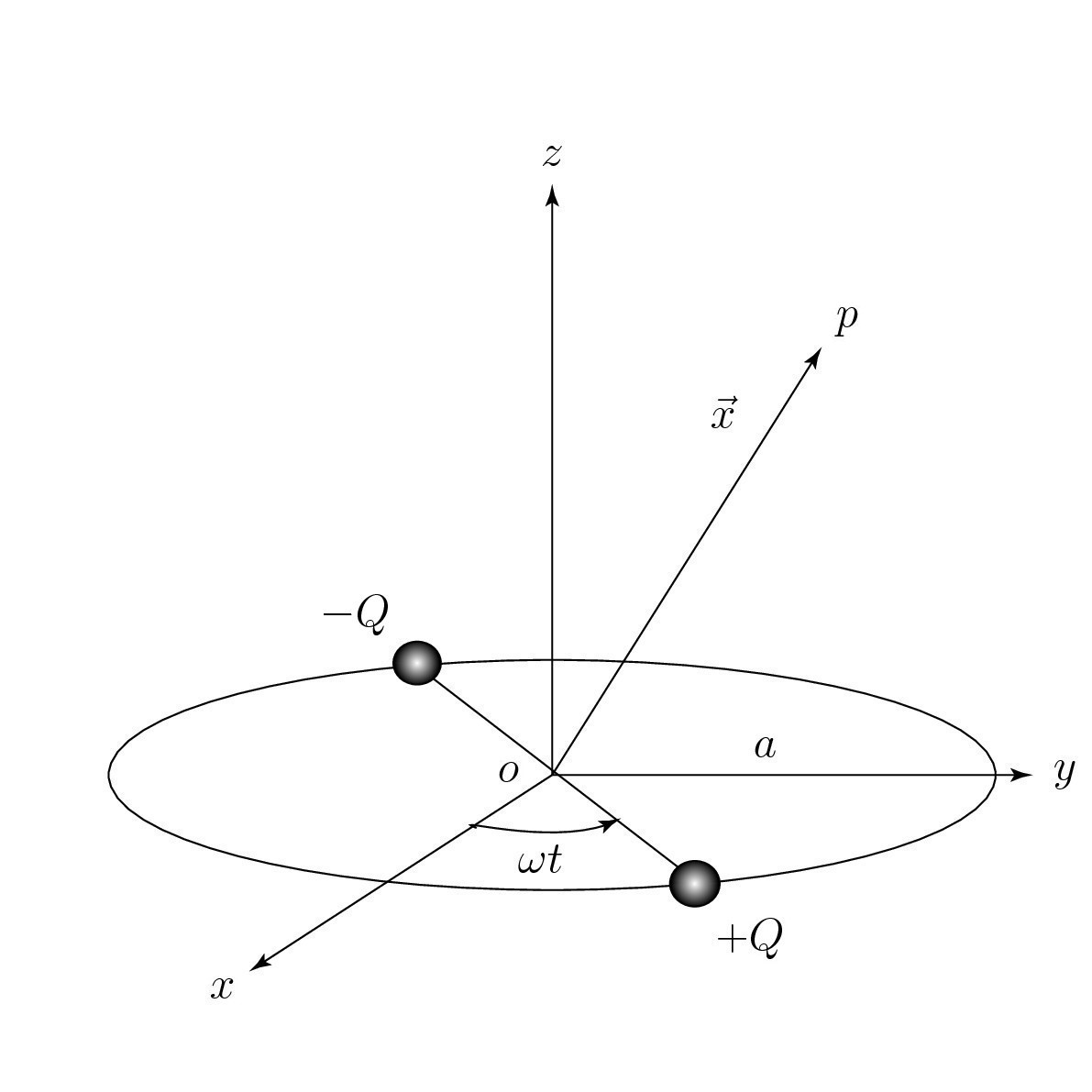}
\vskip -0.6cm
\caption[gw3] {The schematic diagram for a rotating charge electric dipole moment consisting of charges $+Q$ and $-Q$ which 
co-rotate in circular orbits of radius $a$, with an angular frequency $\omega$.}
\label{gw3}
\end{figure}
%%%%%%%%%%%%%%%%%%%%%%%%%%%%%%%%%%%%%%%%%%%%%%%%%%%%%%%%%%%%%%%%%%%%%%%%%%%%%

\acknowledgments{The author would like to thank the anonymous referees for helpful comments. He 
was supported by Basic Science Research Program through the National Research Foundation of Korea 
funded by the Ministry of Education, NRF-2019R1I1A1A01058449.}\\ \\

%%%%%%%%%%%%%%%%%%%%%%%%%%%%%%%%%%%%%%%%%%%%%%%%%%%%%%%%%%%%%%%%%%%%%%%%%
\appendix
\section{EM wave radiation from rotating charge electric dipole moment}\label{chargeedm}
\setcounter{equation}{0}
\renewcommand{\theequation}{A.\arabic{equation}}
%%%%%%%%%%%%%%%%%%%%%%%%%%%%%%%%%%%%%%%%%%%%%%%%%%%%%%%%%%%%%%%%%%%%%%%%%%

In this appendix we will pedagogically investigate the charge electric dipole moment~\cite{jackson99,baylis02,griffiths99}
which possesses charges $+Q$ and $-Q$ rotating on the $x$-$y$ plane. 
Here the charge electric dipole moment is not merging to yield no distortion. 
We assume that for simplicity the two charges co-rotate in circular orbits of radius $a$ about their common center of mass $o$ 
with an angular frequency $\omega$, as shown in Figure 3. Note that the rotating charge electric dipole can be thought of as the 
superposition of two oscillating charge electric dipoles, one along the $x$-axis and the other along the $y$-axis~\cite{griffiths99}. 
The rotating charge vector $\vec{Q}(t)$ on the $x$-$y$ plane is given by
\beq
\vec{Q}(t)=Q(\hat{x}\cos\omega t+\hat{y}\sin\omega t).
\label{qtvector}
\eeq
For the charges on the $x$-axis and $y$-axis, exploiting (\ref{qtvector}) we obtain 
the Green retarded charge scalar potential $\phi_{Q}^{I}$ and $\phi_{Q}^{II}$, respectively
\bea
\phi_{Q}^{I}(\vec{x},t)&=&\frac{1}{4\pi\epsilon_{0}}\left(\frac{Q\cos[\omega(t-R_{+}/c)]}{R_{+}}
-\frac{Q\cos[\omega(t-R_{-}/c)]}{R_{-}}\right),\nn\\
\phi_{Q}^{II}(\vec{x},t)&=&\frac{1}{4\pi\epsilon_{0}}\left(\frac{Q\sin[\omega(t-\bar{R}_{+}/c)]}{\bar{R}_{+}}
-\frac{Q\sin[\omega(t-\bar{R}_{-}/c)]}{\bar{R}_{-}}\right),
\label{qretpot10}
\eea 
where $R_{\pm}$ and $\bar{R}_{\pm}$ are given by (\ref{rpm}). Here the subscript $Q$ stands for the charge 
related quantities. 
Following the technical procedures similar to those in Section 3, we find the total contributions to the 
charge scalar potentials $\phi_{Q}=\phi_{Q}^{I}+\phi_{Q}^{II}$
\beq
\phi_{Q}(\vec{x},t)=-\frac{Q\omega a}{2\pi\epsilon_{0}}\left(\frac{\sin\theta}{cr}\right)
\left(\cos\phi\sin\left[\omega\left(t-\frac{r}{c}\right)\right]
-\sin\phi\cos\left[\omega\left(t-\frac{r}{c}\right)\right]\right),
\label{qretpottotal}
\eeq 
where we have used approximations $a \ll r$ and $\frac{1}{r}\ll\frac{\omega}{c}$, implying that 
the wavelength $\lambda\left(=\frac{2\pi c}{\omega}\right)$ of the EM wave is much smaller than $r$ to yield $\lambda\ll r$.

Making use of (\ref{qtvector}) for the charges co-rotating on the $x$-$y$ plane, we find the charge current 
\beq
\vec{I}_{Q}(t)=-Q\omega(\hat{x}\sin \omega t-\hat{y}\cos\omega t).
\label{qimtcurr}
\eeq 
Next we find the retarded charge vector potentials $\vec{A}_{Q}^{I}$ and 
$\vec{A}_{Q}^{II}$ for the charge currents on the $x$-axis and the $y$-axis, respectively
\bea
\vec{A}_{Q}^{I}(\vec{x},t)&=&\frac{\mu_{0}Q\omega}{4\pi}
\left(-\int_{-a}^{0}\frac{\sin[\omega(t-R_{-}^{\prime}/c)]}{R_{-}^{\prime}}dx^{\prime}
-\int_{0}^{a}\frac{\sin[\omega(t-R_{+}^{\prime}/c)]}{R_{+}^{\prime}}dx^{\prime}\right)\hat{x},\nn\\
\vec{A}_{Q}^{II}(\vec{x},t)&=&
\frac{\mu_{0}Q\omega}{4\pi}
\left(\int_{-a}^{0}\frac{\cos[\omega(t-\bar{R}_{-}^{\prime}/c)]}{\bar{R}_{-}^{\prime}}dy^{\prime}
+\int_{0}^{a}\frac{\cos[\omega(t-\bar{R}_{+}^{\prime}/c)]}{\bar{R}_{+}^{\prime}}dy^{\prime}\right)\hat{y},
\label{qam1}
\eea 
where $R_{\pm}^{\prime}$ and $\bar{R}_{\pm}^{\prime}$ are given by (\ref{rpmprime22}). 
Here we have used the approximations $a \ll r$ and $\frac{1}{r}\ll\frac{\omega}{c}$. 
Note that in (\ref{qretpot10}) and (\ref{qam1}) we have used the charges $(+Q,-Q)$, 
distinct form the binary system of compact objects where we have exploited the masses $(M,M)$ in (\ref{retpot10}) and (\ref{am1}). 
Exploiting the charge vector potentials in (\ref{qam1}), we obtain the total retarded charge vector potential 
$\vec{A}_{Q}=\vec{A}_{Q}^{I}+\vec{A}_{Q}^{II}$
\bea
\vec{A}_{Q}(\vec{x},t)&=&\frac{\mu_{0}Q\omega a}{2\pi r}\left[-\left(\sin\left[\omega\left(t-\frac{r}{c}\right)\right]
+\frac{\omega a}{2c}\sin\theta\cos\phi\cos\left[\omega\left(t-\frac{r}{c}\right)\right]\right)\hat{x},\right.\nn\\
&&\left.+\left(\cos\left[\omega\left(t-\frac{r}{c}\right)\right]-\frac{\omega a}{2c}\sin\theta\sin\phi
\sin\left[\omega\left(t-\frac{r}{c}\right)\right]\right)\hat{y}\right].
\label{qamtotal}
\eea

Now using (\ref{qretpottotal}) and (\ref{qamtotal}), we find the charge electric field 
$\vec{E}_{Q}\equiv  -\na\phi_{Q}-\frac{\pa \vec{A}_{Q}}{\pa t}=E_{Q}^{\theta}\hat{\theta}
+E_{Q}^{\phi}\hat{\phi}$ and charge magnetic one $\vec{B}_{Q}\equiv \na\times\vec{A}_{Q}=B_{Q}^{\theta}\hat{\theta}+B_{Q}^{\phi}\hat{\phi}$
\bea
E_{Q}^{\theta}&=&cB_{Q}^{\phi}=\frac{\mu_{0}Q\omega^{2} a}{2\pi r}\cos\theta\left(\cos\left[\omega\left(t-\frac{r}{c}\right)\right]\cos\phi
+\sin\left[\omega\left(t-\frac{r}{c}\right)\right]\sin\phi\right),\nn\\
E_{Q}^{\phi}&=&-cB_{Q}^{\theta}=\frac{\mu_{0}Q\omega^{2} a}{2\pi r}\left(-\cos\left[\omega\left(t-\frac{r}{c}\right)\right]\sin\phi
+\sin\left[\omega\left(t-\frac{r}{c}\right)\right]\cos\phi\right),
\label{qeb}
\eea 
%-------------------to insert
where we have included the relevant leading terms. From (\ref{qeb}), we obtain $\vec{B}_{Q}=\frac{1}{c}(\hat{r}\times\vec{E}_{Q})$. 
Note that $\vec{E}_{Q}$ does not have the $E_{Q}^{r}$ component, different form the $\vec{E}_{M}$ of 
the binary compact objects which possesses the $E_{M}^{r}$ component in (\ref{eb}).

%-------------------to insert
Next we formulate the charge Poyinting vector in terms of $\vec{E}_{Q}$ and $\vec{B}_{Q}$ in (\ref{qeb}) 
\beq
\vec{S}_{Q}\equiv\frac{1}{\mu_{0}}(\vec{E}_{Q}\times\vec{B}_{Q})
=\frac{1}{\mu_{0}c}(E_{Q})^{2}\hat{r}
\label{qpoynting}
\eeq
where 
\beq
(E_{Q})^{2}=(E_{Q}^{\theta})^{2}+(E_{Q}^{\phi})^{2}=\left(\frac{\mu_{0}Q\omega^{2} a}{2\pi r}\right)^{2}\left(1-\sin^{2}\theta\cos^{2}
\left[\omega\left(t-\frac{r}{c}\right)-\phi\right]\right).
\label{qpoynting2}
\eeq
Now, using (\ref{qpoynting}) and (\ref{qpoynting2}) and averaging in time over a complete cycle, we arrive at 
the EM radiation intensity, namely the EM radiation power per surface $\Sigma$
\beq
\frac{dP_{Q}}{d\Sigma}\equiv \langle\vec{S}_{Q}\rangle=\frac{1}{\mu_{0}c}\left(\frac{\mu_{0}Q\omega^{2} a}{2\pi r}\right)^{2}
\left(1-\frac{1}{2}\sin^{2}\theta
\right)\hat{r}.
\label{qpoynting3}
\eeq
The total EM radiation power constructed by integrating $\langle\vec{S}_{Q}\rangle$ over a sphere $\Sigma_{r}$ of radius $r$ is then given by
\beq
P_{Q}\equiv\int\langle\vec{S}_{Q}\rangle\cdot d\vec{\sigma}=\frac{2\mu_{0}Q^{2}\omega^{4}a^{2}}{3\pi c},
\label{qpoynting4}
\eeq
where $d\vec{\sigma}$ is an area element vector perpendicular to the surface $\Sigma_{r}$. 
Note that the total EM radiation power $P_{Q}$ in (\ref{qpoynting4}) for the rotating charge electric dipole moment 
on the $x$-$y$ plane is twice that for the oscillating charge electric dipole 
moment along the $x$-direction, assuming the same time dependence possessing an angular frequency $\omega$ 
of these rotating and oscillating cases.

%%%%%%%%%%%%%%%%%%%%%%%%%%%%%%%%%%%%%%%%%%%%%%%%%%%%%%%%%%%%%%%%%%%%%%%%%
%\appendix
\section{TT gauge for rotating binary compact objects}\label{ttgauge}
\setcounter{equation}{0}
\renewcommand{\theequation}{B.\arabic{equation}}
%%%%%%%%%%%%%%%%%%%%%%%%%%%%%%%%%%%%%%%%%%%%%%%%%%%%%%%%%%%%%%%%%%%%%%%%%%

In this appendix we will address brief comments on the TT gauge~\cite{wald84,hobson06,carroll04} for wave 
radiation from rotating binary system related with (\ref{standardlgr}). To accomplish this, we first introduce the TT gauge which is 
defined by choosing~\cite{hobson06}  
\beq
\bar{h}_{TT}^{0i}\equiv 0~~~{\rm and}~~~\bar{h}_{TT}\equiv 0.
\label{htt000i}
\eeq
Here the second condition in (\ref{htt000i}) implies that $\bar{h}_{TT}^{\alpha\beta}=h_{TT}^{\alpha\beta}$ in (\ref{anotherh2}).  

Next we study the space-space component of $\bar{h}_{TT}^{ij}$, which is given by the integrated stress within the source. Note that 
this stress may be described in terms of the quadrupole moment in the TT gauge. Now we define the quadrupole moment tensor
for the source as follows~\cite{wald84,hobson06,carroll04}
\beq
I^{ij}(t)=\int T^{00}x^{i}x^{j}d^{3}x,
\label{iij}
\eeq
where $x^{i}=(x,y,z)$. Together with the relation $T^{00}=c^{2}\rho_{M}(\vec{x},t)$, (\ref{iij}) produces
\beq
I^{ij}(t)=c^{2}\int \rho_{M}(\vec{x},t)x^{i}x^{j}d^{3}x.
\label{iij2}
\eeq
We assume for simplicity that the mass orbits reside on the $x$-$y$ plane as shown in Figure 1(a). 
Exploiting the geometry in Figure 1(a), we read off the coordinates of $\vec{x}(M_{1})$ and $\vec{x}(M_{2})$:
\beq
\vec{x}(M_{1})=(a\cos\omega t, a\sin\omega t,0),~~~\vec{x}(M_{2})=(-a\cos\omega t, -a\sin\omega t,0),
\label{coordm1m2}
\eeq
to yield the mass density
\beq
\rho_{M}(\vec{x},t)=M[\delta(x^{1}-a\cos\omega t)\delta(x^{2}-a\sin\omega t)+\delta(x^{1}+a\cos\omega t)\delta(x^{2}+a\sin\omega t)]\delta(x^{3}).
\label{rhoexp}
\eeq
Inserting $\rho_{M}(\vec{x},t)$ in (\ref{rhoexp}) into (\ref{iij2}) we arrive at 
\beq
I^{ij}(t)=Mc^{2}a^{2}\left(
\begin{array}{ccc}
1+\cos (2\omega t) &\sin (2\omega t) &0\\
\sin (2\omega t)  &1+\cos (2\omega t) &0\\
0 &0 &0
\end{array}
\right).
\label{iijt}
\eeq
Next we find $\bar{h}_{TT}^{ij}$ in terms of the quadrupole moment tensor $I^{ij}(t)$~\cite{wald84,hobson06,carroll04}
\beq
\bar{h}_{TT}^{ij}=-\frac{2G}{c^{6}r}\frac{d^{2}I^{ij}(t)}{dt^{2}}|_{retarded},
\label{httij}
\eeq
where the derivative is evaluated at the retarded time $t^{\prime}=t-r/c$. Combining the quadrupole moment tensor in (\ref{iijt}) and $\bar{h}_{TT}^{ij}$ in 
(\ref{httij}) we end up with
\beq
\bar{h}_{TT}^{ij}=\frac{8GMa^{2}\omega^{2}}{c^{4}r}\left(
\begin{array}{ccc}
\cos [2\omega (t-r/c)] &\sin [2\omega (t-r/c)] &0\\
\sin [2\omega (t-r/c)]  &-\cos [2\omega (t-r/c)] &0\\
0 &0 &0
\end{array}
\right),~~~(i,j=1,2,3).
\label{httij2}
\eeq
which reproduces $\bar{h}_{TT}^{ij}$ in (\ref{standardlgr}).

Now we study the time-time component of $\bar{h}_{TT}^{00}$. To do this we first consider the one-body object of mass $M$ 
located at the position $o$ in Figure 1(a). We then find the potential $\bar{h}^{00}$ as follows
\beq
\bar{h}^{00}=\frac{4GM}{c^{2}r}.
\label{hbar00}
\eeq
For the radiative part $\bar{h}_{TT}^{00}$ of the two-body system of interest, we find~\cite{hobson06}
\beq
\bar{h}_{TT}^{00}=0,
\label{hbar00tt}
\eeq
which, together with (\ref{htt000i}) and (\ref{httij2}), yields (\ref{standardlgr}). 
Note that the first condition in the TT gauge in (\ref{htt000i}) cannot explain the DOF of the non-vanishing mass current density 
associated with the mass velocity $v^{i}$ $(\neq 0)$ and 
the ensuing $J^{i}_{M}\equiv \rho_{M}v^{i}$ which is explicitly included in the Wald approximation in the MLGR in the 
gravitational wave radiation phenomenology in Sections 2 and 3.

%%%%%%%%%%%%%%%%%%%%%%%%%%%%%%%%%%%%%%%%%%%%%%%%%%%%%%%%%%%%%%%%%%%%%%%%%
%\appendix
\section{Formalism of massive spin-one graviton}
\label{spinone}
\setcounter{equation}{0}
\renewcommand{\theequation}{C.\arabic{equation}}
%%%%%%%%%%%%%%%%%%%%%%%%%%%%%%%%%%%%%%%%%%%%%%%%%%%%%%%%%%%%%%%%%%%%%%%%%%

In this appendix we will study the formulation of a {\it massive} graviton in the relativistic quantum mechanics (RQM). 
Note that the photon has been known to be massive in the stringy photon model~\cite{hong211}. Note also that, in the massless limit, the graviton has the same 
physical and mathematical structures as the photon as shown in (\ref{squarebarh5}). Since the formulation of the Hamiltonian in 
the RQM for a massive photon has been well-established~\cite{hong22qm}, we will follow this algorithm to investigate 
the massive graviton. To do this, for simplicity we assume 
that the graviton trajectory is a straight line along 
the $z$ direction. Exploiting $E^{2}=m^{2}+p^{2}$ where 
$p=p_{z}=|\vec{p}|$, we find the relativistic equation of motion for the massive graviton given by
\footnote{Here $\phi^{a}$ stands for the wave function for the massive graviton, explicitly given by 
$\phi^{a}_{A}=(\phi_{1}^{a},\phi_{2}^{a})^{T}$ where the superscript $T$ denotes the transpose of the wave function components. 
Here the spin index $a$ $(a=0,1,2,3)$ 
denotes the spin DOF for the massive graviton with spin one. The component index $A$ $(A=1,2)$ stands for the two DOF which have the 
same DOF of the positive and negative energy solutions with the energy index $\pm$ in (\ref{phipm}), since 
the positive and negative energy solutions are given by linear combinations of the two wave functions with the component 
indices. Note that the wave function $\phi^{a}_{A}$ is described in terms of a $1\times(2_{energy}\otimes 4_{spin})=1\times 8$ column vector. 
From now on, for simplicity we will drop the index $A$ in the wave functions except (\ref{vectorj}).}
\beq
H\phi^{a}=i\partial_{0}\phi^{a}.
\label{hpsi}
\eeq

Now we include a possibility of a negative energy solution for the massive graviton in this work, by following the formalism for the massive photon. 
For the corresponding Hamiltonian for the massive graviton, we introduce 
a minimal $2\times 2$ matrix associated with the positive and negative energy solutions. Next,
following the RQM for the massive photon, we 
proceed to find $H$ in (\ref{hpsi}) for the massive graviton. 
The Hamiltonian $H$ is then given by a $2\times 2$ matrix acting on the component index $A$ only
\beq
H=\vec{\cal A}\cdot\vec{p}+{\cal B}m,
\label{hamiltonian}
\eeq
where ${\cal A}_{i}$ ($i=1,2,3$) and ${\cal B}$ are $2\times 2$ matrices. Making use of $E^{2}=m^{2}+p^{2}$, we obtain the algebra between 
${\cal A}_{i}$ and ${\cal B}$:
\beq
\{{\cal A}_{i},{\cal A}_{j}\}=2\delta_{ij}I,~~~{\cal B}^{2}=I,~~~\{{\cal A}_{i},{\cal B}\}=0,~~~
{\cal A}_{i}^{\dagger}={\cal A}_{i},~~~{\cal B}^{\dagger}={\cal B},
\label{aibeta}
\eeq
where $I$ is a $2\times 2$ unit matrix. Note that
eigenvalues of ${\cal A}_{i}$ or ${\cal B}$ are $\pm 1$ and ${\rm tr}{\cal A}_{i}={\rm tr}{\cal B}=0$.

As in the RQM for the massive photon, exploiting the above relations in (\ref{aibeta}) and 
the massive graviton Hamiltonian in (\ref{hamiltonian}), we find the representations for ${\cal A}_{i}$ and ${\cal B}$ given by 
${\cal A}_{1}={\cal A}_{2}=0$, ${\cal A}_{3}=\sigma_{1}$ and ${\cal B}=\sigma_{3}$, where $\sigma_{i}$ are the Pauli matrices. Inserting the 
above representations for ${\cal A}_{i}$ and ${\cal B}$ into (\ref{hamiltonian}), we obtain the $2\times 2$ Hamiltonian:
\beq
H=\left(
\begin{array}{cc}
m &-i\partial_{3}\\
-i\partial_{3} &-m 
\end{array}
\right).
\label{hpsi2}
\eeq
Making use of (\ref{hpsi}) and (\ref{hpsi2}), we find the relativistic equation of motion for the massive graviton
\beq 
(i\Gamma^{\mu}\pa_{\mu}-m)\phi^{a}(x)=0,
\label{brqm2}
\eeq
where $\phi^{a}(x)$ is a function of $x^{\mu}$ and $\Gamma^{\mu}$ is given by $\Gamma^{\mu}=(\sigma_{3}, 0, 0, i\sigma_{2})$.

Now we investigate the phenomenological aspects of the RQM for the massive graviton. To do this, we start with 
the equation in (\ref{brqm2}). Since, for the relativistic massive graviton 
satisfying the relation $E^{2}=m^{2}+p^{2}$, we have two kinds of solutions corresponding to $E=\pm (m^{2}+p^{2})^{1/2}\equiv \pm p_{0}$, 
we introduce an ansatz for the wave function $\phi^{a}$:
\beq
\phi^{a}(x)=\phi^{a}(p^{\mu})e^{\mp ip_{\sigma}x^{\sigma}},
\label{phipmu}
\eeq 
for the positive and negative solutions with the upper and lower signs, respectively.

Next we find the positive energy solution $\phi_{+}^{a}(x)$ with $E>0$ 
and the negative energy solution $\phi_{-}^{a}(x)$ with $E=-|E|<0$, respectively, 
\beq 
\phi_{+}^{a}(x)=u^{a}(p^{\mu})e^{-ip_{\sigma}x^{\sigma}},~~~
\phi_{-}^{a}(x)=v^{a}(p^{\mu})e^{+ip_{\sigma}x^{\sigma}},
\label{phipm}\eeq
where $u^{a}(p^{\mu})$ and $v^{a}(p^{\mu})$ are given by
\beq 
u^{a}(p^{\mu})=\epsilon^{a}\left(\frac{E+m}{2m}\right)^{1/2}\left(\begin{array}{c}
1\\
\frac{p}{E+m}
\end{array}\right),~~~
v^{a}(p^{\mu})=\epsilon^{a}\left(\frac{|E|+m}{2m}\right)^{1/2}\left(\begin{array}{c}
\frac{p}{|E|+m}\\
1
\end{array}\right).
\label{negsol}
\eeq
Here $\epsilon^{a}$ is a unit polarization vector having the spacetime index $a$ $(a=0,1,2,3)$ 
which is the same as the spin index and is needed to incorporate minimally the spin DOF for the massive graviton. Here we have 
considered the Lorentz frame where $\epsilon^{a}$ is purely space-like so that we can readily find that 
$\epsilon^{a}\epsilon^{a}=\vec{\epsilon}\cdot\vec{\epsilon}=1$, or 
$\epsilon_{a}\epsilon^{a}=-\vec{\epsilon}\cdot\vec{\epsilon}=-1$. Next we have the relation $p_{a}\epsilon^{a}\neq 0$,
\footnote{Here we 
use the units of $\hbar=1$ to yield $p^{\mu}=\hbar k^{\mu}=k^{\mu}$ and $p_{\mu}\epsilon^{\mu}=k_{\mu}\epsilon^{\mu}\neq 0$ as 
in (\ref{kalphaneq}).} since for the massive graviton we have longitudinal component in addition to transverse ones, 
similar to the phonon associated with massive particle lattice 
vibrations~\cite{phonon}. In the RQM for the photon, we have the photon 
and anti-photon corresponding to the positive and negative energy solutions, respectively. Similar 
to the RQM for the photon, we have the massive graviton with negative energy solution.

Reshuffling the equation of motion in (\ref{brqm2}) we obtain with $\bar{\phi}^{a}\equiv\phi^{a\dagger}\Gamma^{0}$
\beq
i\sigma_{3}\pa_{0}\phi^{a}-\sigma_{2}\pa_{3}\phi^{a}-m\phi^{a}=0,~~~
i\pa_{0}\bar{\phi}^{a}\sigma_{3}-\pa_{3}\bar{\phi}^{a}\sigma_{2}+m\bar{\phi}^{a}=0,
\label{eomsigma1}
\eeq 
with which we find the probability continuity equation $\pa_{0}\rho+\na\cdot\vec{J}=0$, where the probability density $\rho$ and probability current $\vec{J}$ 
are given by
\beq
\rho\equiv\bar{\phi}^{a}\sigma_{3}\phi^{a}=\phi^{a\dagger}\phi^{a}=\phi_{1}^{*}\phi_{1}+\phi_{2}^{*}\phi_{2},~~~
\vec{J}\equiv\bar{\phi}^{a}(i\sigma_{2})\phi^{a}\hat{z}=\phi^{a\dagger}\sigma_{1}\phi^{a}\hat{z}=(\phi_{1}^{*}\phi_{2}
+\phi_{2}^{*}\phi_{1})\hat{z}.
\label{vectorj}
\eeq
Here we have used the notation $\phi^{a}\equiv\epsilon^{a}(\phi_{1},\phi_{2})^{T}$. Note that the probability density $\rho$ is positive 
definite in the RQM for the massive graviton. For the positive energy solution with $E>0$, inserting $\phi^{a}_{+}(x)$ in (\ref{phipm}) into 
$\rho$ and $\vec{J}$ in (\ref{vectorj}), we obtain $\rho=\frac{E}{m}$ and $\vec{J}=\frac{\vec{p}}{m}$. 
For the negative energy solution with $E=-|E|<0$ where 
$|E|=(m^{2}+p^{2})^{1/2}$, inserting $\phi^{a}_{-}(x)$ in (\ref{phipm}) into $\rho$ and $\vec{J}$, we find $\rho=\frac{|E|}{m}$ and 
$\vec{J}=\frac{\vec{p}}{m}$.

It seems appropriate to comment on the negative energy solution $\phi_{-}^{a}$ in the RQM for the massive graviton. 
First, for the negative energy solution of the massive graviton, since we have $\rho=|\phi_{1}|^{2}+|\phi_{2}|^{2}>0$, $\rho=\frac{|E|}{m}$ implies that $m$ is positive. Even in this negative energy solution, the positive mass $m$ moves with the probability current $\vec{J}$ along the $z$ direction. 
From now on, we will name the graviton possessing the properties that the particle has the positive mass $m$ and 
positive energy $|E|$ and is associated with the negative energy solution, an {\it anti-graviton}.

Second, we propose that the anti-graviton related with the negative energy solution is defined to interact repulsively with the ordinary massive graviton, oppositely to the ordinary massive graviton-graviton attractive gravitational interaction pattern. Note that, in the Dirac RQM, the positron associated with the negative energy solution is defined to interact attractively with the electron, oppositely to the ordinary charged electron-electron repulsive 
electromagnetic interaction pattern. The same logic can be applied to the gravitational interaction case related with the anti-graviton. 
Next, the positron and electron can annihilate each other via the 
particle and anti-particle pair annihilation mechanism. In contrast, the uncharged anti-graviton scatters away from the ordinary massive particle including the graviton and photon, in the repulsive gravitational interaction between the anti-graviton and the ordinary massive particle.

Third, since the anti-graviton is repulsive against charged or uncharged ordinary massive matters, the anti-graviton does not adhere to the 
ordinary massive matters, so that the anti-gravitons can produce an intense radiation flare of the GRB-like graviton. 
Phenomenologically this anti-graviton could be a candidate for an intense radiation flare of the GRB170817A related with the binary compact 
objects merger event GW170817.

\end{document}